\documentclass[a4paper,11pt]{article}
\usepackage[numbers,sort&compress]{natbib}
\usepackage{jcappub} % for details on the use of the package, please
\usepackage[T1]{fontenc} % if needed

\usepackage{latexsym}
\usepackage{graphicx}
\usepackage{amsmath}
\usepackage{amssymb}
\usepackage{amsfonts}
\usepackage{times}
\usepackage{xspace} 
\usepackage{url}

\usepackage{subfigure}		% for subfigures
\usepackage[normalem]{ulem} % for strikeout

\usepackage{hyperref} % for hyperlinks

\allowdisplaybreaks

\title{\boldmath Stealth black hole perturbations in kinetic gravity braiding}

\author{Reginald Christian Bernardo and Ian Vega}
\affiliation{National Institute of Physics, University of the Philippines, Diliman, Quezon City 1101, Philippines}

\emailAdd{rbernardo@nip.upd.edu.ph}
\emailAdd{ivega@nip.upd.edu.ph}
\abstract{We study stealth black hole perturbations in shift symmetric kinetic gravity braiding and obtain its analogous Regge-Wheeler and Zerilli master equations for the odd and even parity sectors. We show that the nontrivial hair of static and spherically symmetric stealth black holes contributes only an additional source term to the even parity master equation. Furthermore, we derive exact solutions to the monopolar and dipolar perturbations and show that they are generally pathological non-gauge modes, or else reduce to the pure-gauge low-order multipoles of general relativity.}

\begin{document}
\maketitle
\flushbottom

%--------------------------------------------------
% the main text of your paper begins here
%--------------------------------------------------
\section{Introduction}
\label{sec:intro}	

General relativity (GR) is undoubtedly the best scientific theory of gravity we have at present. The observation of gravitational waves (GW) has only solidified this position, confirming GR's prediction that GWs propagate at the speed of light, with only little wiggle room for error \cite{ligo_catalog, ligo_catalog_2}. However, a plethora of alternative theories of gravity are also compatible with luminally-propagating GWs and remain worthy of further examination \cite{gw_170817_ligo, dark_energy_creminelli, dark_energy_ezquiaga, lombriser2016breaking, lombriser2017challenges}. 

Among the many alternatives to GR, scalar-tensor theories are still regarded the most compelling because of theoretical parsimony--they require only one extra field and can be elegantly described by a small number of arbitrary potentials tunable for phenomenological purposes \cite{alternative_gravity_koyama, alternative_gravity_joyce, alternative_gravity_clifton, st_horndeski_seminal, st_galileon_inflation_kobayashi_2, st_horndeski_galileons_charmousis}. From within scalar-tensor theories, kinetic gravity braiding (KGB) \cite{st_galileon_inflation_deffayet, st_galileon_inflation_kobayashi}, described by a set of second-order field equations and two free potentials, has stood out in the last two years because of its compatibility with the existing GW speed constraint and its desirable cosmological features such as scaling and self-tuning mechanisms \cite{scaling_kgb_frusciante, st_horndeski_scaling_solutions_albuquerque, st_horndeski_cubic_appleby, st_horndeski_cosmology_emond2018, self_tuning_bh_emond}. The nonlinear scalar field self-interaction in KGB is also characteristic of the existence of a screening mechanism \cite{screening_mcmanus}, guaranteeing agreement with Solar system tests, and notably two of its limits, namely, the Galileon ghost condensate \cite{peirone2019cosmological, st_dark_energy_tsujikawa, horndeski_review_kobayashi2019} and the generalized cubic covariant Galileon \cite{gccg_frusciante, st_horndeski_cubic_babichev, st_horndeski_symmetry_babichev}, have recently been shown to be compatible with cosmological observations. But for any alternative theory to legitimately compete with GR it must be compatible with all observational tests of gravity. In light of recent and forthcoming breakthroughs in GW astronomy, it is therefore desirable to also look at the black holes of an alternative theory.

The strong gravity regime, epitomized by black holes, is a promising theoretical laboratory for locking-in on observational signatures of a dark degree of freedom (d.o.f.). However, it has been shown that special sectors of scalar-tensor theories, including KGB, can accommodate \textit{stealth} black holes that pretend to be the GR black holes except that they carry an invisible nontrivial scalar field or ``hair'' at the background level \cite{st_stealth_minamitsuji_2014, st_stealth_silva, stealth_ben_achour, stealth_motohashi, st_dhost_bhs_minamitsuji, st_kerr_bhs_charmousis, st_horndeski_solutions_babichev_0, st_horndeski_cosmological_tuning_babichev, stability_babichev, stealth_minamitsuji, stealth_bernardo}. The existence of stealth black holes potentially makes discriminating between GR and alternative theories even more observationally challenging and thus demands investigation at its perturbative regime. This paper is a step in this direction, as we study stealth black hole perturbations in KGB.

Black hole perturbations in scalar-tensor theories have been studied previously using either covariant perturbation theory or effective field theory. In the standard perturbation theory, one starts with a covariant theory, i.e., specified by an action or field equations, and performs a perturbative expansion of the dynamical variables on the black hole background. The linear perturbations about static and spherically symmetric black holes in scalar-tensor theories with second-order field equations have been analyzed in this way \cite{perturbation_kobayashi_1, perturbation_kobayashi_2, st_bh_pt_takahashi} and have for instance led to the intriguing conclusion that the scalar modes on stealth black holes are strongly coupled \cite{stealth_minamitsuji, st_bh_pt_de_rham}. On the other hand, in the effective field theory approach, one analyzes the perturbations in a theory-agnostic fashion by building the most general Lagrangian containing the desired number of d.o.f.s and satisfying the symmetries of a specified background. Isospectrality breaking, the mixing of scalar and tensor modes, and parity violation on static and spherically symmetric black holes have been analyzed in this way \cite{perturbations_eft_tattersall, perturbations_eft_franciolini}. The overlap between the two approaches, however, remains to be examined in detail. We resort to covariant perturbation theory in examining the stealth black holes perturbations in KGB, as this is the natural step starting from a covariant gravitational action (Eq. \eqref{eq:kgb}). 

The perturbations about stealth Kerr black holes have also been recently studied in scalar-tensor theories but without the braiding term \cite{dhost_langlois, st_kerr_pt_charmousis}. The Teukolsky equation was instead only modified by an effective source term from the scalar hair but the modes, however, continue to be strongly coupled. In this paper, we shall develop the Regge-Wheeler formalism to study the \textit{odd} and \textit{even} parity perturbations of nonrotating stealth black holes with a static and spherically symmetric scalar hair and end up with a similar conclusion for its metric radiative modes. We shall also obtain the contribution of the hair to the monopolar and dipolar modes. Interestingly, we find the scalar perturbation to be non-dynamical in stealth black hole backgrounds.

We emphasize that the main new results of this paper -- the existence of a source-modified Zerilli equation (Eq. \eqref{eq:zerilli_k_essence}), nondynamicality of the scalar, and the physical (non-gauge) divergence of the low-order multipoles -- can be directly traced to strong coupling in the stealth sector. Our work completes the exploration of stealth black hole perturbations in KGB, and it hints at a more general connection between the aforementioned peculiar properties of the perturbations and strong coupling. This work is also the first time that analogues to \emph{both} Regge-Wheeler and Zerilli master equations for the odd and even-parity sectors of stealth black hole perturbations have been presented for a broad class of scalar-tensor theories. We view this as an important technical achievement that, together with the detailed derivation we lay bare here, should pave the way to similar derivations in other theories.

The outline of this paper is as follows. We start with a brief overview of KGB and its observational constraints (Section \ref{sec:kgb}). We obtain the linearized field equations of shift symmetric KGB (Section \ref{sec:linearized_perturbations}) and, for both simplicity and concreteness, focus on perturbations on a known stealth anti-de-Sitter black hole in a subset of KGB known as $k$-essence (Section \ref{sec:master_equations_k_essence}). Afterwards, we present the general analysis of perturbations of nonrotating stealth black holes with static and spherically symmetric hair in KGB (Section \ref{sec:perturbations_kgb}). We summarize the relevant conclusions and discuss the issue of strong coupling (Section \ref{sec:discussion}). In the Appendix, we write down the full linearized field equations in KGB (Appendix \ref{sec:linear_expressions}) and the explicit form of the coefficients in static and spherically symmetric backgrounds (Appendix \ref{sec:coefficients_hairy}). We also discuss in detail the restrictive class of $tt$-only monopolar and dipolar gauge modes (Appendix \ref{sec:tt_only_gauge_modes}) and the propagation cones of KGB (Appendix \ref{sec:cones_kgb}).

\textit{Conventions.} We work with the mostly plus metric signature $\left(-, +, +, +\right)$ and geometrized units $\left( c = G = 1 \right)$ where $c$ is the speed of light in vacuum and $G$ is Newton's gravitational constant. For notational simplicity, we use the same symbol for the time- and frequency-domain versions of an observable, e.g., $q\left( t \right) \sim q\left( \omega \right) e^{-i \omega t}$, and warn only whenever we think the distinction must be clear. Coordinates on the two-sphere $\left( \theta, \varphi \right)$ are denoted by uppercase latin indices $A, B, \cdots$. For brevity, the mode labels $(l, m)$ of spherical harmonics are suppressed and the summation over modes and two-sphere indices are implicit, e.g., $f_A\left( x, \theta ,\varphi \right) = \sum_{lm} \sum_{B} t_{lm} \left( x \right) \epsilon_A^{\ B} \partial_B Y_{lm} \left( \theta, \varphi \right) = t \left( x \right) \epsilon_A^{\ B} \partial_B Y_{lm} \left( \theta, \varphi \right) $. Readers are encouraged to download the Mathematica notebooks used for this paper from the authors' GitHub repository \footnote{\textit{\href{https://github.com/reggiebernardo/notebooks}{github.com/reggiebernardo/notebooks}}}.

\section{Kinetic gravity braiding}
\label{sec:kgb}

Kinetic gravity braiding is described by the gravitational action \cite{st_galileon_inflation_deffayet, st_galileon_inflation_kobayashi}
\begin{equation}
\label{eq:kgb}
S_g = \int d^4 x \sqrt{-g} \left[ \kappa R + K \left( \phi, X \right) - G \left( \phi, X \right) \Box \phi \right]
\end{equation}
where $g_{ab}$ is the metric, $R$ is the Ricci scalar, $\kappa = M_P^2/2 = 1/ 16 \pi G$, $\phi$ is the scalar field, $X = - g^{ab} \nabla_a \phi \nabla_b \phi / 2$ is the scalar field's kinetic density, and $K$ and $G$ are arbitrary functions that we refer to as the $k$-essence \cite{kessence_seminal_armendariz, kessence_seminal_armendariz2} and braiding potentials, respectively \footnote{The term ``braiding'' refers to the mixing of scalar ($\psi$) and tensor ($h$) modes present in the term $G \ , \Box \phi \sim G \ , \partial h \ , \partial \psi + O\left( h^2, \psi^2 \right)$ in the second order action \cite{st_galileon_inflation_deffayet}.}
. For brevity, we write down $\xi_a =\nabla_a \xi$, $\xi^a = \nabla^a \xi$, and $\xi_{ab} = \nabla_b \nabla_a \xi$ for any scalar function $\xi$, e.g., $X = -g^{ab} \phi_b \phi_b/2$.

Several astrophysical and cosmological constraints are worth mentioning to make the case of KGB as a competitive theory of gravity. First, KGB belongs to the few theories favored by the GW speed constraint \cite{gw_170817_ligo, dark_energy_creminelli, dark_energy_ezquiaga, lombriser2016breaking, lombriser2017challenges, dark_energy_derham, lisa_prospects_barausse}. Second, subsets of KGB, such as the Galileon ghost condensate \cite{peirone2019cosmological, st_dark_energy_tsujikawa, horndeski_review_kobayashi2019} and the generalized cubic covariant Galileon \cite{gccg_frusciante} have been proven to be just as cosmologically viable as $\Lambda$CDM even with large data sets that have ruled out others \cite{st_horndeski_galileon_barreira_1, st_horndeski_galileon_barreira_2,st_horndeski_renk, st_horndeski_galileon_peirone}. Third, KGB is singled out to be among very special observational limits of cosmologically-modified gravity and is known to keep a screening mechanism to pass Solar system tests \cite{no_slip_linder, no_slip_cmb_brush, no_run_linder, limited_modified_gravity, st_horndeski_vainshtein_dima, kgb_vainshtein_anson}.

For the rest of this work we shall work in the shift symmetric KGB, i.e., $K = K\left(X\right)$ and $G = G\left(X\right)$, where the particular sector accommodating stealth black holes has been discussed in Ref. \cite{stealth_bernardo}. The background field equations of shift symmetric KGB are
\begin{equation}
\label{eq:bg_mfe} 
\begin{split}
\kappa G_{ab} &- \frac{1}{2} g_{ab} K - \frac{1}{2} \phi_a \phi_b K_{X} \\
&+ \left[ \frac{1}{2} \phi_a \phi_b \Box \phi - \phi_{( a} \phi_{b) c} \phi^c + \frac{1}{2} g_{ab} \phi^c \phi^d \phi_{cd} \right] G_X = 8 \pi T^{(\text{M})}_{ab}
\end{split}
\end{equation}
and
\begin{equation}
\begin{split}
\label{eq:bg_sfe} 
\Box \phi K_{X} &- \phi^a \phi^b \phi_{ab} K_{XX} \\
& + \left[ - \phi^a \nabla_a \Box \phi - \left( \Box \phi \right)^2 + \phi^a \Box \phi_a + \phi_{ab} \phi^{ab} \right] G_X \\
& \phantom{avengersassembli} + \phi^b \phi_{ab} \left[ \phi^a \Box \phi - \phi^d \phi_d^{\ a} \right] G_{XX} = 0 
\end{split}
\end{equation}
where a subscript $X$ in the potentials denote differentiation with respect to $X$, e.g., $K_{XX} = d^2 K/ dX^2$, and the symmetrization rule for a tensor $T_{ab}$ is $T_{(ab)} = \left( T_{ab} + T_{ba} \right) /2$. Also, $T^{(\text{M})}_{ab}$ is the matter's stress-energy tensor. Eq. \eqref{eq:bg_mfe} can be regarded as the Einstein equation with an additional scalar field stress-energy tensor given by
\begin{equation}
\begin{split}
8 \pi T^{(\phi)}_{ab} = g_{ab} \frac{K}{2} + \phi_a \phi_b \frac{K_{X}}{2} - \bigg[ \phi_a \phi_b \Box \phi - 2 \phi_{( a} \phi_{b) c} \phi^c + g_{ab} \phi^c \phi^d \phi_{cd} \bigg] \frac{ G_X }{2} .
\end{split}
\end{equation}
On the other hand, Eq. (\ref{eq:bg_sfe}) is the field equation for the scalar. 
A background solution $(g_{ab}, \phi)$ to the scalar-tensor theory $(K, G)$ is one in which Eqs. \eqref{eq:bg_mfe} and \eqref{eq:bg_sfe} are simultaneously satisfied. 

\section{Linearized perturbations in shift symmetric KGB}
\label{sec:linearized_perturbations}

In this section, we present the linearized gravitational field equations of KGB, setting the stage for the derivation of the master equation for nonrotating stealth black holes in the next sections.

Consider the metric and scalar perturbations $( h_{ab}, \psi )$, i.e., 
\begin{eqnarray}
g_{ab} &\rightarrow& g_{ab} + h_{ab} \\
\phi &\rightarrow& \phi + \psi .  
\end{eqnarray}
By expanding the field equations (Eqs. \eqref{eq:bg_mfe} and \eqref{eq:bg_sfe}) about the fields $( h_{ab}, \psi )$ up to first order we obtain the linearized field equations of KGB given by
\begin{equation}
\label{eq:mfe_linear}
\begin{split}
\kappa \delta G_{ab} \left[ h_{cd} \right] - 8\pi \delta T^{(\phi, K)}_{ab} \left[ h_{cd} , \psi \right] - 8\pi \delta T^{(\phi, G)}_{ab} \left[ h_{cd} \right] - 8\pi \delta T^{(\phi, G)}_{ab} \left[ \psi \right]= 8\pi  \delta T_{ab} \left[ \text{M} \right]
\end{split}
\end{equation}
and
\begin{equation}
\label{eq:sfe_linear}
\delta S^{(K)} \left[ h_{cd} , \psi \right] + \delta S^{(G)} \left[ \psi \right] + \delta S^{(G)} \left[ h_{cd} \right] = 0 
\end{equation}
where the functionals $\delta G_{ab}\left[h_{cd}\right]$ (Eq. \eqref{eq:einstein_tensor_linear}), $\delta T^{(\phi, K)}_{ab} \left[ h_{cd} , \psi \right]$ (Eq. \eqref{eq:set_phi_K}), $\delta T^{(\phi, G)}_{ab} \left[ h_{cd} \right]$ (Eq. \eqref{eq:set_phi_G_h}), $\delta T^{(\phi, G)}_{ab} \left[ \psi \right]$ (Eq. \eqref{eq:set_phi_G_psi}), $\delta S^{(K)} \left[ \psi \right]$ (Eq. \eqref{eq:sfe_K_linear}), $\delta S^{(G)} \left[ \psi \right]$ (Eq. \eqref{eq:sfe_G_psi_linear}), and $\delta S^{(G)} \left[ \psi \right]$ (Eq. \eqref{eq:sfe_G_h_linear}) are explicitly shown in Appendix \ref{sec:linear_expressions}. In Eq. \eqref{eq:mfe_linear}, $\delta G_{ab}\left[h_{cd}\right]$ is the well-known expression for the linearized Einstein tensor and $\delta T_{ab} \left[ \text{M} \right]$ is the matter perturbation's stress-energy tensor. The rest of the terms come from the scalar field sector of the theory. Superscripts of $K$ and $G$ stand for the $k$-essence ($K$-dependent) and braiding ($G$-dependent) terms, respectively, in the field equations. In Eqs. \eqref{eq:mfe_linear} and \eqref{eq:sfe_linear}, we further break down the $G$-dependent pieces into $\psi$- and $h_{ab}$-dependent terms as both expressions combined is quite long.

Admitedly, the linearized field equations are rather unwieldy to deal with. They are nonetheless more tractable after specializing to stealth black holes. In particular, the generic scalar field contribution can always be written down as
\begin{equation}
\label{eq:generic_scalar_contrib}
S_{ab} = A_{ab} F_X +  \left( B_{ab} + \phi_{cd}\phi^c C^{d}_{\ ab} \right) F_{XX} + \left( D_{ab} + \phi_{cd}\phi^c E^{d}_{\ ab} \right) F_{XXX}
\end{equation}
where $F$ stands for either of the potentials ($K$ or $G$) and the tensors $A$, $B$, $C$, $D$, and $E$ are functionals of the perturbations $h_{ab}$ and $\psi$. On stealth black holes, the conditions $F_X = 0$ and $\phi_{ab} \phi^b = 0$ are satisfied (see Eqs. \eqref{eq:condition_kx} and \eqref{eq:condition_gx}) and Eq. \eqref{eq:generic_scalar_contrib} drastically simplifies to
\begin{equation}
\label{eq:generic_scalar_contrib_2}
S_{ab} = B_{ab} F_{XX} + D_{ab} F_{XXX} .
\end{equation}
The contributions of the scalar field to the linearized metric and scalar field equations therefore eventually reduce to Eqs. \eqref{eq:set_phi_K_stealth}, \eqref{eq:set_phi_G_stealth}, \eqref{eq:sfe_K_linear_stealth}, \eqref{eq:sfe_G_psi_linear_stealth}, and \eqref{eq:sfe_G_h_linear_stealth}. We'll see this work out in a particular theory (Section \ref{sec:master_equations_k_essence}) and the general case (Section \ref{sec:perturbations_kgb}).

\section{Master equations for gravitational perturbations of a hairy black hole in $k$-essence}
\label{sec:master_equations_k_essence}

Going further requires choosing a background on which the perturbations propagate. Such a hairy black hole solution is presented in Ref. \cite{hairy_bhs_tfl} and for this section we focus on describing its perturbations. We start by presenting the hairy black hole (Section \ref{subsec:tfl_bh}) and reducing the linearized field equations down to component level (Section \ref{subsec:pt_hairy}). In Sections \ref{subsec:odd_hairy} and \ref{subsec:even_hairy}, we decompose the perturbations in spherical harmonics and present the master equation for the odd and even parity sectors.

\subsection{Hairy black hole in $k$-essence theory}
\label{subsec:tfl_bh}

In Ref. \cite{hairy_bhs_tfl}, it was shown that hairy black holes in theories including Eq. \eqref{eq:kgb} cannot acquire cosmological relevance. Nonetheless, it introduced analytical black hole solutions that can be used in other studies. One such solution is a stealth black hole to be described below and will be used in this section to study stealth black hole perturbations.
The hairy black hole described by
\begin{eqnarray}
\label{eq:line_element_hairy} ds^2 &=& - f(r) dt^2 + \frac{dr^2}{f(r)} + r^2 d\Omega^2  \\
\label{eq:metric_hairy} f(r) &=& 1 - \frac{2M}{r} + \frac{\beta^2 r^2}{6\kappa} \\
\label{eq:scalar_hair} \phi^{\prime 2}(r) &=& 2 \beta^2 /f(r)
\end{eqnarray}
is an exact solution to $k$-essence with the potential
\begin{equation}
\label{eq:k_potential_tfl}
K(X) = X + 2 \beta \sqrt{-X} 
\end{equation}
where $\beta$, $\kappa$, and $M$ are constants \cite{hairy_bhs_tfl}. It is easy to show that this is a solution by substituting it back into Eqs. \eqref{eq:bg_mfe} and \eqref{eq:bg_sfe}. The nontrivial scalar field profile, i.e., $\phi' \neq 0$ outside the black hole, represents the scalar hair. The spacetime described by this solution is obviously that of a stealth Schwarzschild-anti de Sitter (SAdS) black hole and the background scalar field is nongravitating \footnote{
The solution can be generalized by considering an electromagnetic field $F^{\mu \nu}$, i.e., consider the additional term $S \sim \int d^4 x \sqrt{-g} q^2 F^{\mu \nu} F_{\mu \nu}$ in the action. For the purely electric case, the metric function $f$ of the $k$-essence given by Eq. \eqref{eq:k_potential_tfl} becomes
\begin{equation}
f(r) = 1 - \frac{2M}{r} + \frac{2 q^2}{r^2} + \frac{\beta^2 r^2}{6 \kappa}
\end{equation}
while the scalar field remains as Eq. \eqref{eq:scalar_hair}. This hairy black hole solution is appropriate for studying electrically-charged perturbations.
}. 

It is instructive to point out that the divergence of the scalar field gradient (Eq. \eqref{eq:scalar_hair}) at the event horizon, $f(r) = 0$, does not imply the unphysicality of the solution because neither $\phi$ nor $\phi'$ couples directly with the metric and/or the matter fields. The divergence therefore does not have a straightforward observational consequence and is also common to general stealth black holes, where the one above is only a special case. Nonetheless, the regularity of the black hole solution can instead be inspected by calculating the Noether current arising from shift symmetry \cite{st_horndeski_babichev}: 
\begin{equation}
\label{eq:shift_current_kgb}
J^a\left( x \right) = -\phi^a \left( K_X - \Box \phi G_X \right) - \left( \partial^a X \right) G_X .
\end{equation}
In terms of this field, the scalar field equation (Eq. \eqref{eq:bg_sfe}) can be written as a total divergence, $\nabla_a J^a \left(x\right) = 0$. By substituting Eq. \eqref{eq:k_potential_tfl} and $G_X = 0$ into Eq. \eqref{eq:shift_current_kgb}, this becomes
\begin{equation}
J^a\left( x \right) = -\phi^a \left( 1 - \dfrac{\beta}{\sqrt{-X}} \right) .
\end{equation}
This vanishes everywhere for the stealth black hole solution given by Eqs. \eqref{eq:line_element_hairy}, \eqref{eq:metric_hairy}, and \eqref{eq:scalar_hair}.

\subsection{Linearized field equations}
\label{subsec:pt_hairy}

We present the component-reduced linearized field equations for the perturbations $(h_{ab}, \psi)$ on the hairy black hole presented in Section \ref{subsec:tfl_bh}. We refer the reader to Appendix \ref{sec:coefficients_hairy} for the component form of coefficients appearing in the linearized equations.

After some work, we find that the linearized field equations on the black hole background can be written as
\begin{equation}
\label{eq:mfe_linear_comp}
\begin{split}
\kappa \delta G_{ab} -\dfrac{\beta^2 h_{ab} }{2} - \dfrac{ \psi_c \phi^c \phi_a \phi_b }{4 \beta^2} + \dfrac{ h_{cd} \phi^c \phi^d \phi_a \phi_b }{8 \beta^2}  = 8 \pi \delta T^{(\text{M})}_{ab}
\end{split}
\end{equation}
and
\begin{equation}
\label{eq:sfe_linear_comp}
\begin{split}
\psi_a \phi^a \Box \phi & - h_{bc} \phi_{a}^{\ c}\phi^a \phi^b + \phi^a \psi_{ba} \phi^b + \phi_{ab} \phi^a \psi^b - \dfrac{ h_{ab} \phi^a \phi^b \Box \phi }{2} \\
& - \dfrac{ \phi^a \phi^b \left( \nabla_c h_{ab} \right) \phi^c }{2} - h_{bc} \phi^a \phi^b \phi_{a}^{\ c} - \dfrac{ 3 \psi_a \phi^a \phi^b \phi_{cb} \phi^c }{2 \beta^2} + \dfrac{ 3 h_{cd} \phi^a \phi_{ba} \phi^b \phi^c \phi^d }{4 \beta^2} = 0 .
\end{split}
\end{equation}
In Eq. \eqref{eq:mfe_linear_comp}, the terms coming after $\kappa \delta G_{ab}$ correspond to $\delta T^{(\phi, K)}_{ab} \left[ h_{cd}, \psi \right]$, the scalar field's stress-energy tensor. Also, in Eq. \eqref{eq:sfe_linear_comp}, we have cancelled out an overall factor $1/2 \beta^2$. A noteworthy observation is that the scalar perturbation $\psi$ satisfies a nondynamical equation, i.e., there are no time derivatives in Eq. \eqref{eq:sfe_linear_comp}. The absence of all terms with $\partial_t \psi$ is a consequence of $K_{X} = 0$ and also of contractions with the static and spherically symmetric vector $\phi_a$. Supporting the claim that the scalar field perturbation, $\psi$, is nondynamical, it can be shown that the effective metric of the scalar modes is nonhyperbolic. This implies that the scalar field does not propagate and that its sound speed is infinite \cite{perturbation_kobayashi_1, perturbation_kobayashi_2}. The scalar modes therefore react instantaneously to its source $h_{ab}$ and must be nonradiative.

By explicitly computing all contractions, we further end up with the linearized Einstein equation
\begin{equation}
\label{eq:mfe_linear_2}
\kappa \delta G_{ab} - \frac{\beta^2}{2} h_{ab} - \frac{\phi'}{2} \left( \partial_r \psi \right) \delta_a^{\ r} \delta_b^{\ r} + \frac{\beta^2}{2} h_{rr} \delta_a^{\ r} \delta_b^{\ r} = 8\pi T_{ab}^{(M)} 
\end{equation}
and the linearized scalar field equation
\begin{equation}
\label{eq:sfe_linear_2}
\partial_r^2 \psi + \partial_r \ln \left( \phi^{\prime -2} r^2 \right) \partial_r \psi = \beta^2 \frac{h_{rr}}{\phi'} \partial_r \ln \left( r^2 \phi^{\prime -3} h_{rr} \right) .
\end{equation}
Eq. \eqref{eq:sfe_linear_2} is a linear, first-order, differential equation for the radial gradient, $\partial_r \psi$, with the exact solution
\begin{equation}
\label{eq:scalar_field_grad_1}
\partial_r \psi = \beta^2 \left[ \dfrac{ h_{rr} \left( t, r, \theta, \varphi \right) }{ \phi' \left( r \right) }+ I(t, \theta, \varphi) \frac{\phi^{\prime 2}}{r^2} \right] .
\end{equation}
where $I$ is an integration constant in the coordinate $r$. This is the black hole's scalar fluctuation in a surprisingly bold form. In this result, it must be stressed that $I$ is a perturbation, representing the scalar field fluctuation's d.o.f., or rather what's left of it, after the perturbation equations have been evaluated on the stealth black hole background. New terms sourced by $I$ can therefore be interpreted to come directly from the black hole's scalar hair. It is also worth mentioning that the factor besides $I$ in Eq. \eqref{eq:scalar_field_grad_1} may blow up at the event horizon (see Eq. \eqref{eq:scalar_hair}). This divergence signals a breakdown of perturbation theory at the horizon and can be controlled by setting $I = 0$ or, in other words, removing the scalar hair's influence on the perturbations entirely. However, it must also be mentioned that neither $\psi$ nor its derivative couple directly with the metric and matter perturbations and so the observational consequences of this divergence, if any, requires further study. In what follows, we shall keep the $I$ terms in order to see what it might otherwise contribute to the master equations if it were set to zero. Also, keeping $I$ this way sets the stage for the generalization to KGB in the next section where the calculations are less tractable.

We can bring Eq. \eqref{eq:scalar_field_grad_1} back to the linearized Einstein equation (Eq. \eqref{eq:mfe_linear_2}) and obtain
\begin{equation}
\label{eq:linearized_pert_hairy_bh}
\begin{split}
& \kappa \delta G_{ab}^{(1)} - \frac{\beta^2}{2} h_{ab} + \frac{ \beta^2 }{2} I \left( t, \theta ,\varphi \right) \frac{ \phi^{\prime 3 } \left( r \right) }{ r^2 } \delta_a^r \delta_b^r  = 8 \pi \delta T_{ab}^{(M)} .
\end{split}
\end{equation}
The third term in the left hand side of Eq. \eqref{eq:linearized_pert_hairy_bh} shows that the hair of the black hole modifies only the $rr$-component of linearized field equation for the metric. 

For what it's worth, we discuss the scaling properties of $\partial_r \psi$ (Eq. \eqref{eq:scalar_field_grad_1}) at infinity and the event horizon.
To do so, we express the black hole's scalar fluctuation as
\begin{equation}
\label{eq:scalar_field_grad_1_again}
\partial_r \psi = \beta \sqrt{ \dfrac{f \left( r \right)}{2} } h_{rr} \left( t, r, \theta, \varphi \right) + \dfrac{ 2 \beta^4 }{r^2 f\left( r \right)} I(t, \theta, \varphi) .
\end{equation}
The natural boundary condition at the event horizon is causal, i.e., no radiation exiting from the black hole's interior. This can be deduced from the wave equation for the even parity metric perturbations (Eq. \eqref{eq:zerilli_k_essence}) and imposes $h_{rr} \sim e^{-i \omega \left( t + r_* \right) }$ where $r_*$ is the tortoise coordinate (Eq. \eqref{eq:tortoise}). At the horizon, where $f \left( r \right)$ vanishes, the first term of Eq. \eqref{eq:scalar_field_grad_1_again} then vanishes while the second term blows up. However, at spatial infinity, the boundary condition for an AdS black hole may be reflective but in general it deserves a much deeper discussion. We refer the reader to Refs. \cite{qnms_berti, Andrade:2015gja}. Nonetheless, this only affects the first term of Eq. \eqref{eq:scalar_field_grad_1_again} and the second term, the one sourced by the scalar hair, clearly vanishes. 

\subsection{Odd parity perturbations}
\label{subsec:odd_hairy}

Eq. \eqref{eq:linearized_pert_hairy_bh} shows that the scalar hair does not modify the odd parity sector of the gravitational spectrum. To understand this, we can simply look at the general, i.e., gauge-free, expression for the odd parity metric perturbations \cite{regge_wheeler_classic, zerilli_classic_2, moncrief_classic, perturbations_martel, perturbations_nagar}:
\begin{eqnarray}
h_{tA} &=& h_0 (r) \epsilon_A^{\ B} \partial_B Y_{lm} \left( \theta, \varphi \right) e^{-i \omega t} \\
h_{rA} &=& h_1 (r) \epsilon_A^{\ B} \partial_B Y_{lm} \left( \theta, \varphi \right) e^{-i \omega t} \\
h_{AB} &=& h_2 (r) \epsilon_{(A}^{\ \ C} \nabla_{B)} \nabla_C Y_{lm} (\theta, \varphi) e^{-i \omega t} 
\end{eqnarray}
where $Y_{lm}$ are the spherical harmonics, $A = \left( \theta , \varphi \right)$, $\epsilon_2^{\ 2} = \epsilon_3^{\ 3} = 0$, $\epsilon_2^{\ 3} = -1 / \sin \theta$, and $\epsilon_3^{\ 2} = \sin \theta$, and the sums over the multipole indices $\left( l, m \right)$ and frequency $\omega$ are implicit. These terms in $h_{ab}$ are independent of the $rr$-component and the spherical symmetry of the background keeps it this way. Consequently, this also implies that the odd parity sector is outside of the influence of the scalar hair correction in Eq. \eqref{eq:linearized_pert_hairy_bh}.

The calculation of the master equation in the odd parity sector therefore proceeds exactly as in GR with a cosmological constant and ends up with the Regge-Wheeler equation. A lot of material is available on this (see, for example, Refs. \cite{regge_wheeler_classic, zerilli_classic_2, moncrief_classic, perturbations_martel, perturbations_nagar}). For completeness, we present the Regge-Wheeler equation for the radiative modes ($l \geq 2$) in frequency-domain.
To get to this, we first similarly decompose the matter fields' stress-energy tensor into its odd parity components:
\begin{eqnarray}
T_{tA} &=& t_0 (r) \epsilon_A^{\ B} \partial_B Y_{lm} (\theta, \varphi) e^{-i \omega t} \\
T_{rA} &=& t_1 (r) \epsilon_A^{\ B} \partial_B Y_{lm} (\theta, \varphi) e^{-i \omega t} \\
T_{AB} &=& t_2 (r) \epsilon_{(A}^{\ \ C} \nabla_{B)} \nabla_C Y_{lm} (\theta, \varphi) e^{-i \omega t} .
\end{eqnarray}
Substituting the above odd parity metric and matter perturbations into Eq. \eqref{eq:linearized_pert_hairy_bh}, working in the Regge-Wheeler gauge $h_2(r) = 0$, eliminating $h_0$ using the $\theta\varphi$-component, defining the master function
\begin{equation}
\Psi_{\text{odd}} \left( r \right) = \dfrac{f\left(r\right) h_1 \left(r\right)}{r}, 
\end{equation}
and solving for the $r\varphi$-component, then we obtain Regge-Wheeler equation
\begin{equation}
\label{eq:regge_wheeler_k_essence}
- \partial_{r_*}^2 \Psi_{\text{odd}} + \left( V_{\text{odd}}(r) - \omega^2 \right) \Psi_{\text{odd}} = s_{\text{odd}}
\end{equation}
where the source term and effective potential are 
\begin{equation}
s_{\text{odd}} = - \frac{8\pi f}{\kappa r^3} \left[ 2 r^2 f t_1 + 2 \left( 3M - r \right) t_2 + r^2 \partial_{r*} t_2 \right]
\end{equation}
and 
\begin{equation}
\label{eq:V_odd_tfl}
V_{\text{odd}} = \frac{f}{r^3} \left( l \left( l + 1 \right) r - 6 M \right) ,
\end{equation}
respectively. The tortoise coordinate $r_*$ is defined, in the usual way, by
\begin{equation}
\label{eq:tortoise}
\frac{dr_*}{dr} = 1/f(r) .
\end{equation}
It is worth noting that the special case $t_0 = t_1 = t_2 = 0$ does not necessarily imply vacuum, e.g., for radially-plunging matter orbits, this condition is satisfied and only the even parity sector is excited.

For the odd parity dipole mode ($l = 1$) \footnote{There is no odd parity monopole mode ($l = 0$).}, the odd parity tensor harmonic, $\epsilon_{(A}^{\ \ C} \nabla_{B)} \nabla_C Y_{lm} (\theta, \varphi)$, vanishes and so the available gauge degree of freedom can be used to reduce the number of independent components to just one. This can be easily solved exactly (as in Ref. \cite{zerilli_classic_2}). For concreteness, it can be shown that the odd parity dipole component modifies the $t\varphi$-component of the metric as
$\delta g_{t\varphi} = -2 J \sin^2 \theta / r$ where $J$ is an integration constant. This modification describes the exterior spacetime of a slowly rotating compact object with angular momentum $J$.

\subsection{Even parity perturbations}
\label{subsec:even_hairy}

In this section, we derive the master equation for the even parity modes with $l \geq 2$ of the SAdS black hole of Section \ref{subsec:tfl_bh}. For completeness, we also exactly solve for the monopole ($l = 0$) and dipole ($l = 1$) components and discuss their modifications due to the scalar hair.

We proceed in the Regge-Wheeler gauge \footnote{By looking at Eq. \eqref{eq:linearized_pert_hairy_bh}, it would seem that a natural gauge for the even parity sector is one in which $h_{rr} = 0$. We were unable to obtain a master equation in such gauges.}, i.e.,
\begin{eqnarray}
h_{tt} &=& f(r) H_0 \left( t, r \right) Y_{lm} \left(\theta, \varphi\right) \\
h_{tr} &=& H_1 \left( t, r \right) Y_{lm} \left(\theta, \varphi\right) \\
h_{rr} &=& H_2 \left( t, r \right) Y_{lm} \left(\theta, \varphi\right) / f(r) \\
h_{tA} &=& 0 \\
h_{rA} &=& 0 \\
h_{AB} &=& r^2 K \left( t, r \right) \gamma_{AB} Y_{lm} \left( \theta, \varphi \right) 
\end{eqnarray}
where $\gamma_{AB}$ is the metric on the unit two-sphere. To start, let us restrict our attention to the only place in Eq. \eqref{eq:linearized_pert_hairy_bh} where modifications enter. In frequency-domain, i.e., $h_{ab} \sim e^{-i\omega t}$, the scalar hair contribution to Eq. \eqref{eq:linearized_pert_hairy_bh} is given by
\begin{equation}
\label{eq:hair_rr}
\text{``hair"} =  \frac{q \left( \omega \right) \beta^2}{r^2 f \left( r \right) ^{3/2}}
\end{equation}
where $q \left( \omega \right)$ is the Fourier transform of $\sqrt{2} \beta^3 I \left( t \right)$. Thus, the $rr$-component of the Eq. \eqref{eq:linearized_pert_hairy_bh} becomes 
\begin{equation}
\label{eq:hair_rr_simple}
\text{``$rr$-SAdS"} + \frac{q \left( \omega \right) \beta^2}{r^2 f \left( r \right) ^{3/2}} = \text{``matter''}
\end{equation}
where $\text{``$rr$-SAdS"}$ and ``matter'' correspond to terms coming from the Einstein-Hilbert part of the theory and the matter sector, respectively.

To prepare for the actual derivation of the even parity master equation, we first setup the general even parity matter perturbations
\begin{eqnarray}
T_{tt} &=& f(r) T_0(r) Y_{lm} (\theta, \varphi) e^{-i \omega t} \\
T_{tr} &=& T_1(r) Y_{lm} (\theta, \varphi) e^{-i \omega t} \\
T_{rr} &=& T_2(r) Y_{lm} (\theta, \varphi) e^{-i \omega t} / f(r) \\
T_{tA} &=& t_0 (r) \partial_A Y_{lm} (\theta, \varphi) e^{-i \omega t} \\
T_{rA} &=& t_1 (r) \partial_A Y_{lm} (\theta, \varphi) e^{-i \omega t} \\
T_{AB} &=& r^2 \left( T_K \left( r \right) \gamma_{AB} Y_{lm} \left( \theta, \varphi \right) + T_G \left( r \right) \nabla_A \nabla_B Y_{lm} \left( \theta, \varphi \right) \right) e^{-i \omega t} .
\end{eqnarray}
From this point, we can proceed in the same way as in GR by treating the term $q \left( \omega \right) \beta^2/\left( r^2 f^{3/2} \right)$ as an artificial matter source.
Substituting the even parity metric and matter perturbations into Eq. \eqref{eq:linearized_pert_hairy_bh}, we can first eliminate $H_2$ using the $\theta\varphi$-component. This leaves behind three first-order differential equations for $H_0$, $H_1$, and $K$ given by the $t\theta$-, $tr$-, and $r\theta$-components respectively. These equations are then used to reduce the $rr$-component into an algebraic constraint, $\mathcal{C}\left[ H_0, H_1, K \right] = 0$, to be used for the elimination of $H_0$. Following the footsteps of Zerilli \cite{zerilli_classic}, the two remaining perturbations, $H_1 = \omega \mathcal{R}$ and $K$, are then dealt with by writing down
\begin{eqnarray}
K \left( r \right) &=& f_1 \left( r \right) \hat{K} \left( r \right)+ f_2\left( r \right) \hat{R} \left( r \right) \\
\mathcal{R} \left( r \right) &=& f_3 \left( r \right) \hat{K} \left( r \right)+ f_4 \left( r \right) \hat{R} \left( r \right) .
\end{eqnarray}
By requiring that $\hat{K}$ satisfies a Schr\"{o}dinger-like equation, the coefficients, $\{ f_i \}$, are then determined to be
\begin{eqnarray}
f_1(r) &=& \frac{2 \kappa  \sigma  f(r)}{r \left(3 \kappa  r f'(r)+2 \kappa  \sigma -\beta^2 r^2\right)}+f'(r)-\frac{\beta^2 r}{2 \kappa }+\frac{\sigma }{r} \\
f_2(r) &=& 1 \\
f_3(r) &=& -\frac{i \left(-3 \kappa  r f'(r)+2 \kappa  r f_1(r)-2 \kappa  \sigma +\beta^2 r^2\right)}{2 \kappa  f(r)} \\
f_4(r) &=& - \dfrac{ i r } {f(r)} 
\end{eqnarray}
where $2 \sigma = l(l+1) - 2$. At this point, the linearized equations for $\left( \hat{K}, \hat{R} \right)$ are given by
\begin{equation}
\label{eq:partial_kh}
\partial_{r_*} \hat{K} - \hat{R} = -\frac{8 i \pi  r f \left(2 t_0+r T_1\right)}{\kappa  \omega  (3 M+r \sigma )}
\end{equation}
and
\begin{equation}
\label{eq:partial_rh}
\begin{split}
\partial_{r_*} \hat{R} + \left( \omega^2 - V_{\text{even}(r)} \right) \hat{K} = 
& \frac{8 \pi \kappa  r^2 f(r) \left( r T_2 -2 (3 M+r \sigma ) T_G \right)}{\kappa ^2 r (3 M+r \sigma )} \\
& + \frac{8 \pi  \kappa  r^2 f(r)^2 \left( 2 \omega  (3 M+r \sigma ) t_1+i r \sigma  T_1 \right)}{\kappa ^2 r \omega  (3 M+r \sigma )^2} \\
& + \frac{8 \pi i f(r) t_0 \left(12 \kappa  M^2+M \left(6 \kappa  r \sigma -\beta ^2 r^3\right)+2 \kappa  r^2 \sigma  (\sigma +1)\right)}{\kappa ^2 r \omega  (3 M+r \sigma )^2} \\
& - \dfrac{ q\left(\omega\right) \beta^2 }{\kappa} \dfrac{ \sqrt{f(r)}}{ 3M + r \sigma }.
\end{split}
\end{equation}
In Eq. \eqref{eq:partial_rh}, the potential $V_\text{even}$ is given by
\begin{equation}
\label{eq:v_zerilli}
V_{\text{even}}(r) = f(r) \frac{ 3 M^2 r \left( 6 \kappa \sigma +\beta ^2 r^2\right)+6 \kappa \sigma^2 M r^2+ 2 \kappa \sigma^2 (\sigma+1) r^3+18 \kappa  M^3 }{ \kappa  r^3 \left( \sigma r+3 M\right)^2} .
\end{equation}
By differentiating Eq. \eqref{eq:partial_kh} with respect to $r_*$ and eliminating $\partial_{r_*} \hat{R}$ using Eq. \eqref{eq:partial_rh}, then the Zerilli master equation with $\Psi_{\text{even}}\left(r\right) = \hat{K}\left(r\right)$ can finally be obtained.
This straightforwardly leads to master equation given by
\begin{equation}
\label{eq:zerilli_k_essence}
- \partial_{r_*}^2 \Psi_{\text{even}} + \left( V_{\text{even}}(r) - \omega^2 \right) \Psi_{\text{even}} = \tilde{s}_{\text{even}}
\end{equation}
where
\begin{equation}
\tilde{s}_{\text{even}} = s_{\text{even}} + \frac{\beta ^2 q \left( \omega \right)}{\kappa}  \frac{ \sqrt{f(r)} }{ \sigma r+3 M }
\end{equation}
and $s_{\text{even}}$ is the GR source term:
\begin{equation}
\label{eq:s_gr}
\begin{split}
3 \kappa^2 r \omega  (3 M+r \sigma )^2 \dfrac{ s_{\text{even}} }{8 \pi  f} = 
& + 2i t_0 \bigg( 18 \kappa  M^2+M \left(3 \kappa  r (\sigma -3)-6 \beta ^2 r^3\right) \\
& \phantom{ggggggggggggg} +r^2 \sigma  \left(3 \kappa  (\sigma +1)-\beta ^2 r^2\right) \bigg) \\
& + 6 \kappa  r^2 \omega  f (3 M+r \sigma ) t_1 \\
& + i r \left(18 \kappa  M^2-6 M r \left(\kappa  (\sigma +3)+\beta ^2 r^2\right)-\beta ^2 r^4 \sigma \right) T_1 \\
& - 3 \kappa r^2 \left( 3M + r \sigma \right) \bigg( - r \omega T_2 + 2 \left( 3M + r \sigma \right) \omega T_G \\
& \phantom{ggggggggggggggggggiiiiiiiii} + i \left( 2 \partial_{r*} t_0 + r \partial_{r*} T_1 \right) \bigg) .
\end{split}
\end{equation}
It can be checked that the effective potential for the odd (Eq. \eqref{eq:V_odd_tfl}) and even (Eq. \eqref{eq:v_zerilli}) parity master equations agree with that of a Schwarzschild-AdS black hole \cite{qnms_ads_cardoso, qnms_berti}. The matter perturbation coefficients $\left( T_0, T_1, T_2, t_0, t_1, T_K, T_G \right)$ appearing in the GR source term can be fully determined by specifying the orbits of the matter perturbations, e.g., $\{ T_i = 0, t_i = 0 \}$ in the scalar-tensor vacuum and $t_0 = t_1 = T_K = T_G = 0$ for a purely-radial geodesic. Appendix B of Ref. \cite{perturbations_nagar} provides their explicit forms for point particle perturbations. This detail is of course irrelevant in understanding the main results of this paper.

The main result in the calculation of Eq. \eqref{eq:zerilli_k_essence} is that the scalar hair manifests only as an effective source term
\begin{equation}
\frac{\beta ^2 q \left( \omega \right) }{ \kappa } \frac{ \sqrt{f(r)}}{ \sigma r+3 M }
\end{equation}
to the master equation for the radiative ($l \geq 2$) even parity modes. This is a good place to remind the reader that $q \left( \omega \right)$ entered the calculation through the integration constant $I \left( t \right) \sim q\left(\omega\right) e^{-i \omega t}$ (Eq. \eqref{eq:scalar_field_grad_1}) and it describes the scalar perturbation's d.o.f., or what is left of it, on the stealth black hole. This important result reveals that the scalar fluctuations influence the black hole perturbations only as an additional, \textit{unconstrained}, source term.

We suspect several results from this. First, since the hair of the black hole only enters as an additional source term to the even parity sector, the odd parity quasinormal spectra of the hairy black hole is expected be indistinguishable from that of a SAdS black hole in GR. Second, we can expect to see contribution from the scalar hair to gravitational waveforms, e.g., the modified gravity may have a significant effect to the inspiral and merger phases of a binary. Furthermore, the effective stress-energy tensor for GWs should be the Isaacson stress-energy tensor in GR because the scalar modes are nondynamical (and hence nonradiative). The energy flux, proportional to $\Psi_{\text{odd}}^2+\Psi_{\text{even}}^2$, must still be applicable to the analysis of the orbital decay.

For the monopole ($l = 0$), the vector, $\nabla_A Y_{lm} (\theta, \varphi)$, and tensor, $\nabla_A \nabla_B Y_{lm} (\theta, \varphi)$, harmonics vanish as the spherical harmonics $Y_{lm} (\theta, \varphi)$ is a constant. In this case, the two (instead of three for $l \geq 2$) gauge degrees of freedom can be used to set $H_1 = 0$ and $K = 0$. The linearized equations are exactly solved by
\begin{equation}
\label{eq:H0_monopole}
H_0 = \dfrac{ c_1 }{ r f \left( r \right) } + c_2 - \dfrac{ \beta^2 q\left( \omega \right) }{ \kappa } \int^r \dfrac{ dx }{x f\left( x \right)^{3/2}} 
\end{equation}
and
\begin{equation}
\label{eq:H2_monopole}
H_2 = \dfrac{c_1}{ r f \left( r \right)}
\end{equation}
where $c_1$ and $c_2$ are integration constants. The solution attached to $c_1$ describes a mass shift while the constant $c_2$ is just a gauge mode \cite{zerilli_classic_2}. The scalar hair modification to the monopole is therefore given by
\begin{equation}
\label{eq:monopole_hair_tfl}
\delta H_0 = - \dfrac{ \beta^2 q\left( \omega \right) }{ \kappa } \int^r \dfrac{ dx }{x f\left( x \right)^{3/2}} .
\end{equation}
In Appendix \ref{sec:tt_only_gauge_modes} we prove that only a restrictive class of monopolar gauge modes can support a perturbation with only a $tt$-component. The monopolar perturbation given by Eq. \eqref{eq:monopole_hair_tfl} is therefore not a gauge mode and must have physical consequences. This asymptotically diverges at the event horizon as $\delta H_0 \sim \left( r - r_H \right)^{-1/2}$ unless the lower bound of the integral is calibrated to $r = r_H$. On the other hand, at asymptotic infinity, this monopole modification declines as $\delta H_0 \sim r^{-3}$.

For the dipole ($l = 1$), the tensor $\nabla_A \nabla_B Y_{lm} (\theta, \varphi)$ vanishes identically. The three even parity gauge degrees of freedom can be used to reduce the number of independent perturbation components from six (instead of seven for $l \geq 2$) to three $\left( H_0, H_1, H_2 \right)$. In the GR limit, these components can be solved exactly and shown to be just gauge modes describing a center-of-mass shift. The presence of the hairy source term spoils this interpretation. More concretely, the modification $\delta H_0$ due to the scalar hair's even parity dipole is the solution to
\begin{equation}
\delta H_0 ' \left( r \right) - \dfrac{\delta H_0 \left( r \right)}{r f \left( r \right)} = \dfrac{\beta^2 q\left(\omega\right)}{\kappa r f \left( r \right)^{3/2}} .
\end{equation}
As this is only a first-order differential equation, we can express the exact solution as
\begin{equation}
\label{eq:dipole_hair_tfl}
\begin{split}
\delta H_0 = \dfrac{\beta^2 q\left(\omega\right)}{\kappa} & \exp \left( \int^r \dfrac{dx}{x f\left(x\right)} \right) \times \int^r dy \dfrac{ \exp \left( - \int^y \dfrac{dz}{z f\left(z\right)} \right) }{ y f\left(y\right)^{3/2} } .
\end{split}
\end{equation} 
As in the monopole term, the dipolar perturbation does not correspond to a gauge mode (see proof in Appendix \ref{sec:tt_only_gauge_modes}). It diverges at the event horizon due to the inverse powers of $f \left( r \right)$ inside the integrals unless the lower bound of the integral is set to $r = r_H$. At asymptotic infinity, this dipole modification drops as $\delta H_0 \sim r^{-3}$.

\section{Master equations for gravitational perturbations of nonrotating stealth black holes in shift symmetric KGB}
\label{sec:perturbations_kgb}

In the previous section, we have shown that it is indeed possible to obtain a master equation in closed form. We now generalize this calculation for the perturbations of \textit{all} static and spherically symmetric stealth black holes in shift symmetric KGB. We start by reviewing the constraints on the potentials to accommodate stealth black holes (Section \ref{subsec:stealth_kgb}) and setup the linearized field equations (Section \ref{subsec:linearized_theory_kgb}). In Sections \ref{subsec:odd_kgb} and \ref{subsec:even_kgb} we present the master equation for odd and even parity sectors.

\subsection{Stealth black holes in KGB}
\label{subsec:stealth_kgb}

Shift symmetric KGB with stealth black holes are described by $k$-essence and braiding potentials satisfying the constraints
\begin{eqnarray}
\label{eq:condition_kx} K_X \left( X_0 \right) &=& 0 \\
\label{eq:condition_gx} G_X \left( X_0 \right) &=& 0 
\end{eqnarray}
where $X_0$ is a constant equal to the background scalar field's kinetic density \cite{stealth_bernardo}. In this case, the static and spherically symmetric stealth black hole solution is given by
\begin{eqnarray}
\label{eq:metric_stealth} f(r) &=& 1 - \frac{2M}{r} + \frac{ K\left(X_0\right) r^2}{6\kappa} \\
\label{eq:scalar_stealth} \phi^{\prime 2}(r) &=& - 2 X_0 /f
\end{eqnarray}
where the line element is given by Eq. \eqref{eq:line_element_hairy}. The sign of the limit $K \left(X_0\right)$ of the $k$-essence potential therefore determines the asymptotic behavior of the solution whereas the stealth anti-de Sitter black hole of Sec. \ref{subsec:tfl_bh} is only a special case of the above solution for the choice of $K\left(X \right)$ given by Eq. \eqref{eq:k_potential_tfl} and $G \left(X\right)$ is a constant. For asymptotically flat spacetimes, e.g., Schwarzschild, the additional constraint
\begin{equation}
\label{eq:condition_k} K \left( X_0 \right) = 0
\end{equation}
must be imposed further to the $k$-essence potential.

We will show that analogous Regge-Wheeler and Zerilli master equations can be obtained for \textit{all} stealth black holes in shift symmetric KGB. In the following calculation, it is useful to keep in mind that a covariantly constant kinetic density implies that
\begin{equation}
\label{eq:constant_x_identity}
\phi^b \phi_{ab} = 0 .
\end{equation}
This identity can be obtained by taking the covariant derivative of $X$ and setting the result to zero, i.e., $\nabla_a X = \nabla_a \left( - \phi^b \phi_b / 2 \right) = - \phi^b \phi_{ab}$. Eqs. \eqref{eq:condition_kx}, \eqref{eq:condition_gx}, \eqref{eq:condition_k}, and \eqref{eq:constant_x_identity} will play a significant role in the simplification of the linearized field equations (see last paragraph of Section \ref{sec:linearized_perturbations} containing Eqs. \eqref{eq:generic_scalar_contrib} and \eqref{eq:generic_scalar_contrib_2}).

\subsection{Linearized field equations in KGB}
\label{subsec:linearized_theory_kgb}

The constraints on the potentials given by Eqs. \eqref{eq:condition_kx}, \eqref{eq:condition_gx}, and \eqref{eq:condition_k} and Eq. \eqref{eq:constant_x_identity} drastically reduce the number of terms present in the linearized field equations. The linear perturbations of the $K$-dependent pieces or quadratic sector of the scalar field's SET (Eq. \eqref{eq:set_phi_K}) are given by
\begin{equation}
\label{eq:set_phi_K_stealth}
\begin{split}
- 8\pi \delta T^{(\phi, K)}_{ab} \left[ h_{cd} , \psi \right] = -\frac{1}{2} K h_{ab} 
+ \frac{1}{2} \phi_a \phi_b \psi_c \phi^c K_{XX}
- \frac{1}{4} h_{cd} \phi_a \phi_b \phi^c \phi^d K_{XX} .
\end{split}
\end{equation}
On the other hand, the linear perturbations of the $G$-dependent pieces or cubic sector of the scalar field's SET (Eqs. \eqref{eq:set_phi_G_h} and \eqref{eq:set_phi_G_psi}) are given by
\begin{equation}
\label{eq:set_phi_G_stealth}
\begin{split}
- 8\pi \delta T^{(\phi, G)}_{ab} \left[ h_{cd} , \psi \right] = 
- \frac{1}{2} \phi_a \phi_b \psi_c \phi^c \Box \phi G_{XX}
+ \frac{1}{4} h_{cd} \phi_a \phi_b \phi^c \phi^d \Box \phi G_{XX}  .
\end{split}
\end{equation}
The potentials $K_{XX}$ and $G_{XX}$ are understandably constants owing to the covariantly constant background kinetic density \cite{stealth_bernardo}.
The noteworthy observation above is that all of the nonvanishing terms in the perturbation of the scalar field's stress-energy tensor have the generic structure $\phi_a \phi_b F \left(x \right) = \delta_a^r \delta _b^r \phi^{\prime 2} F\left(x\right)$ where $F\left(x\right)$ is a scalar function. We therefore find that the scalar hair correction to the Einstein equation only enters the through the $rr$-component. This a general result that is valid for the linear perturbations of all static and spherically symmetric stealth black holes in shift symmetric KGB. Moreover, this is the key to unlocking the master equation for the gravitational perturbations in stealth, nonrotating, black holes in shift symmetric KGB (Section \ref{subsec:even_kgb}).

We can also verify that the scalar field does not propagate on a hyperbolic cone by calculating its field equation. The linear perturbations of the $K$-dependent pieces (Eq. \eqref{eq:sfe_K_linear}) of the scalar field equation are given by
\begin{equation}
\label{eq:sfe_K_linear_stealth}
\begin{split}
\delta S^{(K)} \left[ h_{cd}, \psi \right] =
&- \psi_a \phi^a \Box \phi K_{XX} - \phi^a \psi_{ba} \phi^b K_{XX} \\
&+ \frac{1}{2} h_{ab} \phi^a \phi^b \Box \phi K_{XX}
+ \frac{1}{2} \phi^a \phi^b \phi^c \nabla_c h_{ab} K_{XX}  .
\end{split}
\end{equation}
The $G$- and $\psi$-dependent pieces (Eq. \eqref{eq:sfe_G_psi_linear}) of the scalar field equation are given by
\begin{equation}
\label{eq:sfe_G_psi_linear_stealth}
\begin{split}
\delta S^{(G)} \left[ \psi \right] =
& +\psi_a \phi^a \nabla_b \Box \phi \phi^b G_{XX} + \psi_a \phi^a \left( \Box \phi \right)^2 G_{XX} \\
& + \phi^a \psi_{ba} \phi^b \Box \phi G_{XX} - \psi_a \phi^a \phi^b \Box \phi_b G_{XX} - \psi_a \phi^a \phi_{cb} \phi^{cb} G_{XX} 
\end{split}
\end{equation}
and the $G$- and $h_{ab}$-dependent pieces (Eq. \eqref{eq:sfe_G_h_linear}) of the scalar field equation are given by
\begin{equation}
\label{eq:sfe_G_h_linear_stealth}
\begin{split}
\delta S^{(G)} \left[ h_{cd} \right] =
& - \frac{1}{2} h_{bc} \phi^a \phi^b \phi^c \left( \nabla_a \Box \phi \right) G_{XX} - \frac{1}{2} h_{ab} \phi^a \phi^b \left( \Box \phi \right)^2 G_{XX} \\
& - \frac{1}{2} \phi^a \phi^b \phi^c \Box \phi \left( \nabla_c h_{ab} \right) G_{XX}
+ \frac{1}{2} h_{bc} \phi^a \phi^b \phi^c \left( \Box \phi_a \right) G_{XX} \\
& + \frac{1}{2} h_{ab} \phi^a \phi^b \phi_{dc} \phi^{dc} G_{XX} .
\end{split}
\end{equation}
By adding the above results (Eqs. \eqref{eq:sfe_K_linear_stealth}, \eqref{eq:sfe_G_psi_linear_stealth}, and \eqref{eq:sfe_G_h_linear_stealth}) and performing the contractions, it can therefore be seen that the scalar field equation takes on the generic structure
\begin{equation}
\label{eq:sfe_linear_stealth}
a_1 \left( r \right) \partial_r^2 \psi + a_2 \left( r \right) \partial_r \psi = Q \left[ h_{cd} \right]
\end{equation}
where $a_i$ are functions of $r$ and $Q$ is a functional in terms of the metric perturbation $h_{cd}$. Specifically, $a_1$ comes from the second term in Eq. \eqref{eq:sfe_K_linear_stealth} and the third term in Eq. \eqref{eq:sfe_G_psi_linear_stealth} while $a_2$ comes from the rest of the terms in Eqs. \eqref{eq:sfe_K_linear_stealth} and \eqref{eq:sfe_G_psi_linear_stealth}. The functional $Q$ is sourced by all of the terms in Eq. \eqref{eq:sfe_G_h_linear_stealth}.

Two noteworthy implications standout from the above result. First is that Eq. \eqref{eq:sfe_linear_stealth} verifies that the scalar field does not propagate on a causal cone. The sound speed of the scalar modes can be confirmed to be nonetheless infinite and so the scalar field responds instantaneously to its source. To prove this, we simply note that the sound speed $c_s$ of the scalar modes in a static and spherically symmetric background in KGB is given by \cite{perturbation_kobayashi_1, perturbation_kobayashi_2, stealth_minamitsuji}
\begin{equation}
\label{eq:sound_speed_scalar}
c_s^2 = \dfrac{ 2\kappa\Xi \phi^{\prime 2} \left( 2r^2 \Gamma - \Xi \right) - 16 \kappa^2 r^4 \Sigma/h  }{ 2 \left( 4\kappa r + \Xi \phi'\right)^2 \left( P_1 - \kappa \right) }
\end{equation}
where
\begin{equation}
P_1 = 2 \dfrac{ r^2 \kappa^2 }{ 4r\kappa + \Xi \phi' } \dfrac{ d }{ dr } \left( \ln \left( \dfrac{f}{h}  \right) \right) +4 \dfrac{ d  }{ dr  } \left( \dfrac{ \kappa^2 r^2 }{ 4\kappa r + \Xi \phi' } \right)
\end{equation}
\begin{equation}
\begin{split}
\Sigma = X \bigg[ K_X 
+ 2X K_{XX}
- f\phi' \left( \dfrac{4}{r} + \dfrac{f'}{f}  \right) \left( G_X + X G_{XX} \right)  \bigg]
\end{split}
\end{equation}
\begin{equation}
\Xi = -2 r^2 X G_X
\end{equation}
\begin{equation}
\Gamma = -4 X G_X 
\end{equation}
and the line element is parametrized as
\begin{equation}
ds^2 = -h\left( r\right) dt^2 + \dfrac{dr^2}{f \left( r \right) } + r^2 d\Omega^2 .
\end{equation}
On a stealth background, $h = f$, $K_X = G_X = 0$, then $\Xi = \Gamma = 0$, $\Sigma \neq 0$, and $P_1 = \kappa$. The vanishing of the denominator of Eq. \eqref{eq:sound_speed_scalar} therefore leads to $c_s^2 \rightarrow \infty$. The second point regarding Eq. \eqref{eq:sfe_linear_stealth} is that it is a first-order differential equation for $\partial_r \psi$. Its exact solution $\psi\left[h_{cd}\right]$ is given by
\begin{equation}
\begin{split}
\partial_r \psi = & \exp \left(-\int^r dr_3 \dfrac{a_2 \left( r_3 \right) }{a_1\left( r_3 \right)} \right) \times \bigg[ \int^r dr_1 \dfrac{Q\left[ h_{cd}\right] }{a_1\left( r_1 \right) } \exp \left( \int^{r_1} dr_2 \dfrac{ a_2\left( r_2 \right) }{ a_1\left( r_2  \right)} \right) + I \left( t,\theta, \varphi\right)  \bigg]
\end{split}
\end{equation}
where $I \left( t, \theta, \varphi \right)$ is an integration constant. This expresses the scalar field perturbation in closed form as a functional integral in terms of the metric perturbation. By substituting this exact expression for $\partial_r \psi$ into the scalar field's stress-energy tensor, we can therefore expect to write down a linearized Einstein equation that is an integrodifferential equation in the $rr$-component. However, we shall see that we do not need to bother at all with an integrodifferential equation because the integral is equivalent to a drastically simpler algebraic expression. 

\subsection{Odd parity perturbations}
\label{subsec:odd_kgb}

The results of Section \ref{subsec:linearized_theory_kgb} shows that the scalar field's perturbation $\psi$ enters only as an $rr$-component correction to the Einstein equation. This correction is a functional of only $h_{rr}$ and therefore does not associate with the odd parity sector of the gravitational perturbations. 

We conclude that for all static and spherically symmetric stealth black holes in shift symmetric KGB the odd parity sector of the gravitational perturbations is untouched by the scalar hair. It should be noted, however, that this result can be considered unsurprising at the level of linear perturbations which prevents the mixing of an even parity object, e.g., $\psi$, with odd parity terms. The Regge-Wheeler equation (Eq. \eqref{eq:regge_wheeler_k_essence})  is therefore the master equation for the odd parity perturbations. Also, as in Section \ref{subsec:odd_hairy}, the odd parity dipole mode corresponds to an angular momentum perturbation, i.e., $\delta g_{t\varphi} = -2 J \sin^2 \theta / r$ where $J$ is an integration constant \cite{zerilli_classic_2}.

We note that this conclusion has been reached before within the broader class of degenerate higher-order scalar-tensor theories \cite{perturbation_kobayashi_1, st_bh_pt_takahashi, st_bh_pt_de_rham} and from the effective field theory approach \cite{perturbations_eft_franciolini, scordatura_motohashi}. This paper complements the existing literature by directly obtaining the conclusion within shift symmetric KGB and by working with the field equations instead of the second-order action.

\subsection{Even parity perturbations}
\label{subsec:even_kgb}

To obtain the even parity master equation, we begin by writing down the Einstein equation as
\begin{equation}
\label{eq:ee_kgb_stealth}
\kappa \delta G_{ab} - \dfrac{1}{2} K\left( X_0 \right) h_{ab} - \delta_a^r \delta_b^r \times \delta \tau \left[ h_{rr} \right] = 8\pi \delta T_{ab} \left[ \text{M} \right]
\end{equation}
where $\delta \tau \left[ h_{rr} \right]$ is a scalar functional in $h_{rr}$. The term $-K h_{ab}/2$ in the left hand side of Eq. \eqref{eq:ee_kgb_stealth} tunes the cosmological constant and so determines the asympotic behavior of the black hole. But most importantly, because the correction (third term in the left hand side of Eq. \eqref{eq:ee_kgb_stealth}) enters only through the $rr$-component, the Bianchi identity imposes
\begin{equation}
\nabla^a \left( \delta_a^r \delta_b^r \times \delta \tau \left[ h_{rr} \right] \right) = 0 
\end{equation}
which can be solved to obtain
\begin{equation}
\delta \tau \left[ h_{rr} \right] = \dfrac{ q \left( t, \theta, \varphi \right) }{ r^2 f\left(r\right)^{3/2} }.
\end{equation}
This algebraic term which is notably the generalization of Eq. \eqref{eq:hair_rr} is the residue of the scalar hair entering the Einstein equation. For the radiative modes ($l \geq 2$), the master equation can therefore be obtained in the Regge-Wheeler gauge by treating the correction as an additional stress-energy tensor with only an $rr$-component. This eventually leads to Eq. \eqref{eq:zerilli_k_essence} but with the replacement $\beta^2 \rightarrow K\left( X_0 \right)$ in Eqs. \eqref{eq:s_gr} and \eqref{eq:v_zerilli} and the effective source term
\begin{equation}
\tilde{s}_{\text{even}} = s_{\text{even}} + \dfrac{q \left( \omega \right)}{\kappa} \dfrac{ \sqrt{f\left(r\right)} }{\sigma r + 3 M}
\end{equation}
where $q\left( t , \theta, \varphi \right) \sim q\left(\omega\right) Y_{lm} \left( \theta, \varphi \right) e^{-i \omega t}$. This is the main result of the paper: \textit{the scalar hair for all static and spherically symmetric stealth black holes in shift symmetric KGB manifests only as an additional source term}
\begin{equation}
\label{eq:main_result}
\dfrac{q\left(\omega\right)}{\kappa} \dfrac{ \sqrt{ f\left(r\right) } }{\sigma r + 3 M}
\end{equation}
\textit{in the even parity master equation} (Eq. \eqref{eq:zerilli_k_essence}). Echoing the discussion after Eq. \eqref{eq:zerilli_k_essence}, we emphasize that this correction is completely unconstrained and so its usefulness in practice completely rests on whether it can be determined. Otherwise, the inevitable conclusion may just be $q\left(\omega\right) = 0$, implying that the stealth black hole perturbations in KGB are indistinguishable from that of GR.

To complete the results, we discuss the monopole ($l = 0$) and dipole ($l = 1$) perturbations, both of which can be solved exactly as in Section \ref{subsec:even_hairy}. First, for the monopole ($l = 0$), the perturbation due to the scalar hair is given by
\begin{equation}
\label{eq:monopole_hair_stealth}
\delta H_0 = - \dfrac{ q\left( \omega \right) }{ \kappa } \int^r \dfrac{ dx }{x f\left( x \right)^{3/2}} .
\end{equation}
We emphasize that this is not a gauge mode (see Appendix \ref{sec:tt_only_gauge_modes} for proof) and therefore should come with physical consequences. The divergence at the event horizon can be cured by setting the lower bound of the integral to $r = r_H$. At asymptotic infinity, this monopole perturbation declines as $\delta H_0 \sim r^{-3}$ for the anti de Sitter case but diverges as $\left( r - r_{C} \right)^{-1/2}$ at the cosmological horizon $r = r_{C}$ for the de Sitter case. For the Schwarzschild case, the monopolar perturbation modifies the $tt$-component of the metric by $\delta g_{tt} \sim q\left( \omega \right) \left( r - r_H \right)^{1/2}$ for $r \rightarrow r_H$ and by $\delta g_{tt} \sim q\left( \omega \right) \ln r$ for $r \rightarrow \infty$. The divergence at the horizon can be cured by fixing the lower bound of the integral (Eq. \eqref{eq:monopole_hair_stealth}) to $r = r_H$; however, the perturbed Schwarzschild solution logarithmically diverges at infinity. This exact solution to the monopolar mode of stealth black holes can be interpreted to signal the breakdown of perturbation theory.

For the dipole ($l = 1$), the perturbation due to the scalar hair can be written as
\begin{equation}
\label{eq:dipole_hair_stealth}
\begin{split}
\delta H_0 = \dfrac{q\left(\omega\right)}{\kappa} \exp \left( \int^r \dfrac{dx}{x f\left(x\right)} \right) \int^r dy \dfrac{ \exp \left( - \int^y \dfrac{dz}{z f\left(z\right)} \right) }{ y f\left(y\right)^{3/2} } .
\end{split}
\end{equation} 
This dipolar perturbation is not a gauge mode (see Appendix \ref{sec:tt_only_gauge_modes} for proof) and must come with physical consequences. The divergence at the event horizon can be cured by fixing the lower bound of the integral to $r = r_H$. At asymptotic infinity, this dipole modification drops as $\delta H_0 \sim q\left( \omega \right) r^{-3}$ for the anti de Sitter case but  diverges as $\delta H_0 \sim q\left( \omega \right) \left( r - r_C \right)^{-1/2}$ at the cosmological horizon $r = r_C$ for the de Sitter case. For the Schwarzschild case, the dipole perturbation diverges linearly, $\delta H_0 \sim q\left( \omega \right) r$, at asymptotic infinity. The divergence signals a breakdown of perturbation theory.

\section{Discussion and outlook}
\label{sec:discussion}

We have shown that the scalar hair of static and spherically symmetric stealth black holes in shift symmetric kinetic gravity braiding contributes only a source term to the even parity sector of the perturbations. The odd parity sector is therefore unmodified and so is the odd parity quasinormal and power spectrum. We have also obtained analytical expressions for the monopolar and dipolar perturbations.

In Appendix \ref{sec:cones_kgb}, we derive the propagation cones of KGB on a general covariant background and use this independent calculation to strengthen the result that the sound speed of the scalar modes is infinite on stealth black holes. The scalar field therefore becomes nondynamical on a stealth black hole -- a concrete manifestation of strong coupling. This conclusion generally holds and strongly threatens the physicality of stealth black hole solutions in scalar-tensor theories. However, it should be acknowledged that the strong-coupling is also precisely the reason why a Zerilli master equation can be obtained for stealth black holes. This raises an interesting yet more difficult question: ``Does the existence of a Zerilli equation for stealth black hole perturbations imply strong coupling?''. This unfortunately falls outside the scope of this present paper and is left for future work.

It is important to recognize five recent works which have shed light to the issue of strong coupling of perturbations of stealth black holes in scalar-tensor theories \cite{stability_babichev, stealth_minamitsuji, st_bh_pt_de_rham, st_kerr_pt_charmousis, scordatura_motohashi}. In Ref. \cite{stability_babichev}, the monopolar stealth black hole perturbation in a widely-studied non-KGB scalar-tensor theory \footnote{
The gravitational Lagrangian is $L_g = \kappa \left( R - 2 \Lambda \right) + \beta G^{ab} \phi_a \phi_b - \eta X$ where $\kappa, \Lambda, \beta$ and $\eta$ are constants and $R$ and $G_{ab}$ are the Ricci scalar and Einstein tensor, respectively.
}
was analyzed and it was clarified that this mode would be stable in a Schwarzschild black hole for infinite sound speed despite strong coupling. In Ref. \cite{stealth_minamitsuji}, stealth Schwarzschild solutions were obtained for the first time in KGB and the problem of strong coupling in the perturbations of stealth black holes was pointed out. In Ref. \cite{st_bh_pt_de_rham}, strong coupling in nonrotating stealth black holes was shown to persist in the broader context of degenerate higher-order scalar-tensor theories. This puts scalar-tensor theories in a tight spot and shows that strong coupling can be inherent in stealth solutions in scalar-tensor theories. In Ref. \cite{st_kerr_pt_charmousis}, the analysis of perturbations on stealth Kerr black holes in degenerate higher-order scalar-tensor theories led to the conclusion that the perturbations are governed by a Teukolsky equation with an effective source term. In Ref. \cite{scordatura_motohashi}, it was proposed that the strong coupling problem can be resolved by introducing a heavy Ostrogradsky ghost that would not be triggered or rendered observable for observable energy scales. This led to theories -- dubbed \textit{scordatura} degenerate theories -- which relaxes the degeneracy conditions previously imposed in degenerate higher-order scalar-tensor theories to avoid the Ostrogradsky instability.

The results of this paper complement those of Ref. \cite{st_kerr_pt_charmousis} for the nonrotating black hole limit, but with the particular advantage of having analytical expressions to the monopolar and dipolar perturbations from the scalar field. It should also be pointed out that the analysis of Ref. \cite{st_kerr_pt_charmousis} does not include the braiding sector of the theory. Our work shows that its inclusion makes no difference to the over-all result: the hair of stealth black holes contributes an effective source to the master equation for perturbations. For both this paper and Ref. \cite{st_kerr_pt_charmousis}, the effective source term originates from an unconstrained integration constant, i.e., $q\left(\omega\right)$ in Eq. \eqref{eq:main_result}. This means for example that the integration of the master equation (Eq. \eqref{eq:zerilli_k_essence}) cannot proceed without \textit{a priori} selecting an arbitrary $q\left( \omega \right)$. In contrast, the analytical solutions to the monopolar and dipolar perturbations are fully determined. These low-order multipoles diverge unless $q\left( \omega \right) = 0$ for $l = 0, 1$; they are either unphysical or identical to GR's monopole and dipole.

We have also shown that the scalar modes in stealth, static and spherically symmetric, black holes in shift symmetric KGB are nondynamical, i.e., scalar modes do not satisfy a hyperbolic equation and have infinite sound speed. This suggests that the scalar modes are nonradiative and, hence, that gravitational wave fluxes may be computed using simply the stress-energy tensor for the tensor modes, e.g., Isaacson GW stress-energy tensor. Such flux calculations again first require an adequate resolution to the arbitrariness of the integration constant $q\left( \omega \right)$, though none appears natural to us. The scalar perturbation is fully determined by the background fields and boundary conditions, and it appears unable to be a true degree of freedom. A Hamiltonian analysis will fully flesh this out. Nonetheless, this feature leaves us to speculate broader connections with the \textit{cuscuton} \cite{cuscuton_afshordi, cuscuton_afshordi_2, cuscuton_gomes} -- a $k$-essence theory $\left( K\left( \phi, X \right) = \mu^2 \sqrt{ |2X| } \right)$ with infinite sound speed and no propagating excitations in which the effective metric for the perturbations is singular. This question which could lead to an alternative interpretation of the strong coupling problem will be investigated in detail in a different paper.

To the best of our knowledge, this is the first time that Regge-Wheeler and Zerilli master equations for both odd and even parity sectors of hairy black hole perturbations have been reported for a broad class of theories. It remains unclear whether a similar calculation (or some extension) is feasible in other scalar-tensor theories that support stealth black hole solutions. Non-stealth black hole perturbations demand more attention despite the technical barriers one expects in their direction. For stealth black holes, all of the terms with factors of $K_X, G_X$ and $\phi^b \phi_{ab}$ in Eqs. \eqref{eq:set_phi_K}, \eqref{eq:set_phi_G_h}, \eqref{eq:set_phi_G_psi}, \eqref{eq:sfe_K_linear}, \eqref{eq:sfe_G_psi_linear}, and \eqref{eq:sfe_G_h_linear} vanished. This drastic simplification paved the way for the present analysis, but cannot be anticipated for nonstealth black holes, even for static and spherically symmetric ones. We refer the reader to Refs. \cite{st_no_hair_theorem_hui, st_no_hair_benkel, st_no_hair_theorem_sotiriou_1, st_no_hair_theorem_sotiriou_2, st_horndeski_babichev, st_horndeski_solutions_babichev_2, st_horndeski_slow_rotation_bh_maselli, st_horndeski_solutions_kobayashi, st_horndeski_solutions_rinaldi, st_horndeski_solutions_anabalon, st_horndeski_solutions_minamitsuji, st_horndeski_solutions_gaete, antoniou2018evasion, antoniou2018black, bernardo2019hair} for some of the non-stealth black holes in the literature.

% APPENDIXES

\begin{acknowledgments}
The authors are grateful to Eugeny Babichev for pointing out a particularly helpful earlier analysis of stealth black hole perturbations and to Adrian Villanueva and John Celestial for their feedback on an earlier version of the manuscript. The authors acknowledge the use of the package `xAct' \cite{xact} and its derivatives `xPert' \cite{xpert} and `xCoba' \cite{xcoba}. This research is supported by the University of the Philippines Diliman Office of the Vice Chancellor for Research and Development through Project No. 191937 ORG.
\end{acknowledgments}

\appendix

\section{Linear perturbations of the Einstein tensor, scalar field SET, and scalar field equation}
\label{sec:linear_expressions}

In this section, we present explicit functional expressions for the linear perturbations of tensors and other quantities relevant for the discussion. We start with the well-known expression for the linear perturbations of the Einstein tensor
\begin{equation}
\begin{split}
\label{eq:einstein_tensor_linear}
2 \delta G_{ab}\left[h_{cd}\right] = g_{ab} h^{cd} R_{cd} &- h_{ab} R - \nabla_a \nabla_b h + \nabla_c \nabla_a h_b^{\ c} \\
& + \nabla_c \nabla_b h_a^{\ c} - \Box h_{ab} - g_{ab} \nabla_d \nabla_c h^{cd} + g_{ab} \Box h .
\end{split}
\end{equation}
To organize the analysis, we break the scalar's SET into $K$- and $G$-dependent parts. This corresponds to breaking the analysis into quadratic and cubic sectors of Horndeski theory. The linear perturbations of the $K$-dependent pieces or quadratic sector of the scalar's SET are given by
\begin{equation}
\label{eq:set_phi_K}
\begin{split}
- 32\pi \delta T^{(\phi, K)}_{ab} \left[ h_{cd} , \psi \right] = & -2 K h_{ab} - 2 \psi_a \phi_b K_{X} - 2\phi_a \psi_b K_{X} +2 g_{ab} \psi_c \phi^c K_{X} \\
& -  g_{ab} h_{cd} \phi^c \phi^d K_{X} + 2 \phi_a \phi_b \psi_c \phi^c K_{XX} - h_{cd} \phi_a \phi_b \phi^c \phi^d K_{XX} .
\end{split}
\end{equation}
On the other hand, the linear perturbations of the $G$-dependent pieces or cubic sector of the scalar's SET are given by
\begin{equation}
\label{eq:set_phi_G_h}
\allowdisplaybreaks
\begin{split}
- 32 \pi \delta T^{(\phi, G)}_{ab} \left[ h_{cd} \right] = 
& + 2 h_{cd} \phi_a^{\ d} \phi_b \phi^c G_{X} + 2 h_{cd}\phi_a \phi_{b}^{\ d} \phi^c G_{X} + \phi_a \phi_b \left( \nabla_c h \right) \phi^c G_{X} \\
& - 2 \phi_a \phi_b \phi^c \left( \nabla_d h_c^{\ d} \right) G_{X} + \phi_a \left( \nabla_b h_{cd} \right) \phi^c \phi^d G_{X} + \left( \nabla_a h_{cd} \right) \phi_b \phi^c \phi^d G_{X} \\
& + 2 g_{ab} h_{de} \phi_c^{\ e} \phi^c \phi^d G_{X} + 2 h_{ab} \phi^c \phi_{dc} \phi^d G_{X} - 2 h_{cd} \phi_a \phi_b \phi^{dc} G_{X} \\
& - g_{ab} \phi^c \phi^d \left( \nabla_e h_{cd} \right) \phi^e G_{X} - 2 g_{ab} h_{de} \phi^c \phi^d \phi^{e}_{\ c} G_{X} \\
& + h_{cd} \phi_a \phi_b \phi^c \phi^d \left( \Box \phi \right) G_{XX} - h_{de} \phi_{ac} \phi_b \phi^c \phi^d \phi^e G_{XX} \\
& - h_{de} \phi_a \phi_{bc} \phi^c \phi^d \phi^e G_{XX} + g_{ab} h_{ef} \phi^e \phi_{dc} \phi^d \phi^e \phi^f G_{XX}
\end{split}
\end{equation}
and 
\begin{equation}
\label{eq:set_phi_G_psi}
\allowdisplaybreaks
\begin{split}
- 16 \pi \delta T^{(\phi, G)}_{ab} \left[ \psi \right] = 
& + \psi_a \phi_b \left( \Box \phi \right) G_{X} + \phi_a \psi_b \left( \Box \phi \right) G_{X} + \phi_a \phi_b \left( \Box \psi \right) G_{X} \\
& - \psi_{ac} \phi_b \phi^c G_{X} - \phi_{ac} \psi_b \phi^c G_{X} - \psi_a \phi_{bc} \phi^c G_{X} \\
& - \phi_a \psi_{bc} \phi^c G_{X} - \phi_{ac} \phi_b \psi^c G_{X} - \phi_a \phi_{bc} \psi^c G_{X} \\
& + g_{ab} \phi^c \psi_{dc} \phi^d G_{X} + g_{ab} \phi_{cd} \phi^c \psi^d G_{X} + g_{ab} \phi^c \phi_{dc} \psi^d G_{X} \\
& - \phi_a \phi_b \psi_c \phi^c \left( \Box \phi \right) G_{XX} + \phi_{ad} \phi_b \psi_c \phi^c \phi^d G_{XX} \\
& + \phi_a \phi_{bd} \psi_c \phi^c \phi^d G_{XX} - g_{ab} \psi_c \phi^c \phi^d \phi_{ed} \phi^e G_{XX} .
\end{split}
\end{equation}
For the scalar field equation, the linear perturbations of its $K$-dependent pieces are given by
\begin{equation}
\label{eq:sfe_K_linear}
\begin{split}
2 \delta S^{(K)} \left[ h_{cd}, \psi \right] =
& + 2 \left( \Box \psi \right) K_{X} + \left( \nabla_a h \right) \phi^a K_{X} - 2 \phi^a \left( \nabla_b h_a^{\ b} \right) K_{X} \\
& - 2 h_{ab} \phi^{ba} K_{X} - 2 \psi_a \phi^a \left( \Box \phi \right) K_{XX} + 2 h_{bc} \phi_{a}^{\ c}\phi^a \phi^b K_{XX} \\
& - 2 \phi^a \psi_{ba} \phi^b K_{XX} - 2 \phi_{ab} \phi^a \psi^b K_{XX} - 2 \phi^a \phi_{ba} \psi^b K_{XX} \\
& + h_{ab} \phi^a \phi^b \left( \Box \phi \right) K_{XX} + \phi^a \phi^b \left( \nabla_c h_{ab} \right) \phi^c K_{XX} + 2 h_{bc} \phi^a \phi^b \phi_{a}^{\ c} K_{XX} \\
& + 2 \psi_a \phi^a \phi^b \phi_{cb} \phi^c K_{XXX} - h_{cd} \phi^a \phi_{ba} \phi^b \phi^c \phi^d K_{XXX} .
\end{split}
\end{equation}
As for the $G$-dependent pieces of the scalar field equation, we further break down the terms into $h_{ab}$- and $\psi$-dependent pieces for the reason that both expressions combined is exceedingly long. The $G$- and $\psi$-dependent pieces of the scalar field equation is given by
\begin{equation}
\label{eq:sfe_G_psi_linear}
\begin{split}
\delta S^{(G)} \left[ \psi \right] =
& - \left( \nabla_a \Box \psi \right) \phi^a G_{X} - \left( \nabla_a \Box \phi \right) \psi^a G_{X} - 2 \left( \Box \phi \right) \left( \Box \psi \right) G_{X} \\
& + \psi^a \left( \Box \phi_a \right) G_{X} + \phi^a \left( \Box \psi_a \right) G_{X} + 2 \psi_{ba} \phi^{ba} G_{X} \\
& +\psi_a \phi^a \left( \nabla_b \Box \phi \right) \phi^b G_{XX} + \psi_a \phi^a \left( \Box \phi \right)^2 G_{XX} + \phi^a \psi_{ba} \phi^b \left( \Box \phi \right) G_{XX} \\
& + \phi_{ab} \phi^a \psi^b \left( \Box \phi \right) G_{XX} + \phi^a \phi_{ba} \psi^b \left( \Box \phi \right) G_{XX} + \phi^a \phi_{ba} \phi^b \left( \Box\psi \right) G_{XX} \\
& - \psi_a \phi^a \phi^b \left( \Box \phi_b \right) G_{XX} - 2 \phi^a \psi^b \phi_{cb} \phi_{a}^{\ c} G_{XX} - 2 \phi^a \phi^b \psi_{cb} \phi_{a}^{\ c} G_{XX} \\
&- \psi_a \phi^a \phi_{cb} \phi^{cb} G_{XX} - \psi_a \phi^a \phi^b \phi_{cb} \phi^c \left( \Box \phi \right) G_{XXX} + \psi_a \phi^a \phi^b \phi^c \phi_{dc} \phi_{b}^{\ d} G_{XXX} .
\end{split}
\end{equation}
Finally, the $G$- and $h_{ab}$-dependent pieces of the scalar field equation is given by
\begin{equation}
\label{eq:sfe_G_h_linear}
\begin{split}
2 \delta S^{(G)} \left[ h_{cd} \right] =
& + 2 h_{bc} \left( \nabla_a \phi^{cb} \right) \phi^a G_{X} -  \phi_a^{\ b} \phi^a \left( \nabla_b h \right) G_{X} - 2 \left( \nabla_a h \right) \phi^a \Box \phi G_{X} \\
& - \phi^a \left( \nabla_b \nabla_a h \right) \phi^b G_{X} + 2 \phi^a \left( \nabla_b \nabla_c h_{a}^{\ c} \right) \phi^b G_{X} + \phi^a \left( \nabla_b h \right) \phi_{a}^{\ b} G_{X} \\
& - 2 h_{ac} \phi_{b}^{\ c} \phi^{ba} G_{X} + 4 \phi^a \left( \Box \phi \right) \left( \nabla_c h_{a}^{\ c} \right) G_{X} + \phi_{a}^{\ b} \phi^a \left( \nabla_c h_b^{\ c} \right) G_{X} \\
& - 2 \phi^a \phi_a^{\ b} \left( \nabla_c h_{b}^{\ c} \right) G_{X} - \phi^a \phi^b \left( \Box h_{ab} \right) - 2 h_{ab} \phi^a \left( \Box \phi^b \right) G_{X} \\
& - 2 h_{bc} \phi^{ba} \phi_a^{\ c} G_{X} + 2 h_{ac} \phi^a \left( \nabla^c \Box \phi \right) G_{X} + 4 h_{bc} \left( \Box \phi \right) \phi^{cb} G_{X} \\
& + 2 \left( \nabla_a h_{bc} \right) \phi^a \phi^{cb} G_{X} - 4 \phi^a \left( \nabla_c h_{ab} \right) \phi^{cb} G_{X} - 2 h_{bc} \phi^a \left( \nabla^c \phi_a^{\ b} \right) G_{X} \\
& - h_{bc} \left( \nabla_a \Box \phi \right) \phi^a \phi^b \phi^c G_{XX} + 4 h_{bd} \phi^a \phi^b \phi_{c}^{\ d} \phi_a^{\ c} G_{XX} \\
& - 2 \phi^a \phi_{ba} \phi^b \phi^c \left( \nabla_d h_{c}^{\ d} \right) G_{XX} - 2 h_{bc} \phi_a^{\ c} \phi^a \phi^b \Box \phi G_{XX} \\
& - h_{ab} \phi^a \phi^b \left( \Box \phi \right)^2 G_{XX} - \phi^a \phi^b \left( \nabla_c h_{ab} \right) \phi^c \left( \Box \phi \right) G_{XX} \\
& - 2 h_{bc}\phi^a \phi^b \phi_{a}^{\ c} \left( \Box \phi \right) G_{XX} + h_{bc} \phi^a \phi^b \phi^c \left( \Box \phi_a \right) G_{XX} \\
& + 2 \phi^a \phi^b \phi^c \left( \nabla_d h_{bc} \right) \phi_a^{\ d} G_{XX} + 2 h_{cd} \phi^a \phi^b \phi_{a}^{\ c} \phi_b^{\ d} G_{XX} \\
& - 2 h_{cd} \phi^a \phi_{ba} \phi^b \phi^{dc} G_{XX} + h_{ab} \phi^a \phi^b \phi_{dc} \phi^{dc} G_{XX} + \phi^a \phi_{ba} \phi^b \left( \nabla_c h \right) \phi^c G_{XX} \\
& + h_{cd} \phi^a \phi_{ba} \phi^b \phi^c \phi^d \left( \Box \phi \right) G_{XXX} - h_{cd} \phi^a \phi^b \phi^c \phi^d \phi_{eb} \phi_a^{\ e} G_{XXX} .
\end{split}
\end{equation}

\section{Coefficients for the perturbations of a hairy black hole in $k$-essence}
\label{sec:coefficients_hairy}

To simplify the linearized field equations on top of the hairy black hole background (Eqs. (\ref{eq:line_element_hairy}), (\ref{eq:scalar_hair}), and (\ref{eq:metric_hairy})), we can use the following equations which are valid for static and spherically symmetric background:
\begin{equation}
\phi_a \phi_b \psi_c \phi^c = f \phi^{\prime 3} \left( \partial_r \psi \right) \delta_a^r \delta_b^r 
\end{equation}
\begin{equation}
h_{cd} \phi_a \phi_b \phi^c \phi^d = f^2 \phi^{\prime 4} h_{rr} \delta_a^r \delta_b^r
\end{equation}
\begin{equation}
\Box \phi =  f \phi^{\prime \prime} + f' \phi' + \frac{2f}{r} \phi^\prime
\end{equation}
\begin{equation}
\psi_a \phi^a \Box \phi = f \phi' \partial_r \psi \left(  f \phi^{\prime \prime} + f' \phi' + \frac{2f}{r} \phi^\prime  \right)
\end{equation}
\begin{equation}
\phi_{ab} = \phi'' \delta_a^r \delta_b^r - \Gamma^r_{ab} \phi'
\end{equation}
\begin{equation}
h_{bc} \phi_{a}^{\ c}\phi^a \phi^b = h_r^{\ r} \left( f \phi' \right)^2 \left[ \phi'' + \frac{1}{2} \frac{f'}{f} \phi' \right]
\end{equation}
\begin{equation}
\phi^a \psi_{ba} \phi^b = \left( f \phi' \right)^2 \left( \partial^2_r \psi + \frac{1}{2} \frac{f'}{f} \partial_r \psi \right) 
\end{equation}
\begin{equation}
\phi_{ab} \phi^a \psi^b = \left( f \phi' \right) \left[ \phi^{\prime \prime} + \frac{1}{2} \frac{f'}{f} \phi' \right] f \left( \partial_r \psi \right)
\end{equation}
\begin{equation}
h_{ab} \phi^a \phi^b \Box \phi = h_{rr} \left( f \phi' \right)^2 \left( f \phi^{\prime \prime} + f' \phi' + \frac{2f}{r} \phi^\prime \right)
\end{equation}
\begin{equation}
\nabla_c h_{ab} = \partial_c h_{ab} - \Gamma^d_{ac} h_{db} - \Gamma^d_{bc} h_{ad}
\end{equation}
\begin{equation}
\phi^a \phi^b \phi^c \nabla_c h_{ab} = \left( f \phi' \right)^3 \left( \partial_r h_{rr} + \frac{f'}{f} h_{rr} \right) 
\end{equation}
\begin{equation}
h_{bc} \phi^a \phi^b \phi_{a}^{\ c} = h_r^{\ r} \left( f \phi' \right)^2 \left[ \phi'' + \frac{1}{2} \frac{f'}{f} \phi' \right]
\end{equation}
\begin{equation}
\psi_a \phi^a \phi^b \phi_{cb} \phi^c = \left( \partial_r \psi \right) \left( f \phi' \right)^3 \left( \phi'' + \frac{1}{2} \frac{f'}{f} \phi' \right)
\end{equation}
\begin{equation}
h_{cd} \phi^a \phi_{ba} \phi^b \phi^c \phi^d = h_{rr} \left( f \phi' \right)^4 \left( \phi'' + \frac{1}{2} \frac{f'}{f} \phi' \right)
\end{equation}
These are the coefficients of the linearized field equations for any static and spherically symmetric background in $k$-essence theory. In the hairy black hole of Section \ref{subsec:tfl_bh} where
\begin{equation}
K = \beta^2
\end{equation}
\begin{equation}
K_{X} = 0
\end{equation}
\begin{equation}
K_{XX} = - 1/ 2 \beta^2
\end{equation}
\begin{equation}
K_{XXX} = - 3 / 4 \beta^4
\end{equation}
\begin{equation}
\phi_a = \phi' \delta_a^r
\end{equation}
\begin{equation}
\phi^a = f \phi' \delta^a_r
\end{equation}
\begin{equation}
X = - \beta^2
\end{equation}
\begin{equation}
f \phi^{\prime 2} = 2 \beta^2
\end{equation}
\begin{equation}
\phi' = \beta \sqrt{ \frac{2}{f} }
\end{equation}
\begin{equation}
\phi'' + \frac{f'}{2f} \phi' = 0
\end{equation}
\begin{equation}
f \phi'' + f' \phi' + \frac{2f}{r} \phi' = \phi' \left( \frac{f'}{2} + \frac{2f}{r} \right)
\end{equation}
it can be shown that the coefficients in the linearized field equations reduce to
\begin{equation}
\phi_a \phi_b \psi_c \phi^c = 2 \beta^2 \phi' \left( \partial_r \psi \right) \delta_a^r \delta_b^r
\end{equation}
\begin{equation}
h_{cd} \phi_a \phi_b \phi^c \phi^d = 4 \beta^4 h_{rr} \delta_a^r \delta_b^r
\end{equation}
\begin{equation}
\psi_a \phi^a \Box \phi = 2 \beta^2 \partial_r \psi \left( \frac{f'}{2} + \frac{2f}{r}  \right)
\end{equation}
\begin{equation}
h_{bc} \phi_{a}^{\ c}\phi^a \phi^b = 0
\end{equation}
\begin{equation}
\phi^a \psi_{ba} \phi^b = 2 \beta^2 f \left( \partial^2_r \psi + \frac{1}{2} \frac{f'}{f} \partial_r \psi \right)
\end{equation}
\begin{equation}
\phi_{ab} \phi^a \psi^b = 0
\end{equation}
\begin{equation}
h_{ab} \phi^a \phi^b \Box \phi = 2 \beta^2 h_{rr} f \phi' \left( \frac{f'}{2} + \frac{2f}{r} \right)
\end{equation}
\begin{equation}
\phi^a \phi^b \phi^c \nabla_c h_{ab} = 2 \beta^2 f^2 \phi' \left( \partial_r h_{rr} + \frac{f'}{f} h_{rr} \right)
\end{equation}
\begin{equation}
h_{bc} \phi^a \phi^b \phi_{a}^{\ c} = 0
\end{equation}
\begin{equation}
\psi_a \phi^a \phi^b \phi_{cb} \phi^c = 0
\end{equation}
\begin{equation}
h_{cd} \phi^a \phi_{ba} \phi^b \phi^c \phi^d = 0 .
\end{equation}

\section{$tt$-only-gauge modes}
\label{sec:tt_only_gauge_modes}

In this section, we show that only a very restrictive class of monopole ($l = 0$) and even parity-dipole ($l = 1$) metric perturbations with only a nonzero $tt$-component $\left( h_{tt} = f\left( r \right) H_0\left( r \right) Y_{lm} e^{-i\omega t}\right)$ can be gauge modes, i.e., tunable away by a gauge transformation, and that this class excludes the monopolar and dipolar perturbations discovered in Sections \ref{subsec:even_hairy} and \ref{subsec:even_kgb}. 

\subsection{$tt$-only-monopole-gauge mode}
\label{subsec:tt_monopole}

For the monopole, where $l = 0$  the spherical harmonics $Y_{00}$ is a constant, the gauge vector $\xi^a$ can be decomposed as
\begin{equation}
\xi^a = \left( M_0 \left( r\right) Y_{00} e^{-i \omega t}, M_1 \left( r\right) Y_{00} e^{-i \omega t} , 0, 0 \right) .
\end{equation}
The (infinitessimal) gauge or coordinate transformation, $x^a \rightarrow x^a + \xi^a$, perturbs the metric tensor by $\delta h_{ab} = 2\nabla_{(a} \xi_{b)}$, which has the following independent components:
\begin{eqnarray}
\label{eq:tt_mon_gauge} \delta h_{tt} &=& i \left( 2\omega f M_0 + i M_1 f' \right) Y_{00} e^{-i \omega t} \\ 
\label{eq:tr_mon_gauge} \delta h_{tr} &=& - \dfrac{ i \omega M_1 + f^2 M_0' }{f} Y_{00} e^{-i \omega t} \\
\label{eq:rr_mon_gauge} \delta h_{rr} &=& \dfrac{ -M_1 f' + 2 f M_1' }{f^2} Y_{00} e^{- \omega t} \\
\label{eq:theta_mon_gauge} \delta h_{\theta\theta} &=& 2 M_1 r Y_{00} e^{-i\omega t} .
\end{eqnarray}
A perturbation $h_{ab}$ is a gauge mode if there exists a gauge vector $\left( M_0, M_1 \right)$ for which $h_{ab} = \delta h_{ab}$. For example, the monopolar perturbation with $H_0 = H_2 = c / r f\left( r \right)$ is a (mass shift) gauge mode as explicitly investigated by Zerilli in Ref. \cite{zerilli_classic_2}.

We now focus on the $tt$-only-gauge modes. To solve this problem, we must search for the gauge vector $\left( M_0, M_1 \right)$ for which the resulting metric perturbation $\delta h_{ab}$ has only a nonzero $tt$-component. Obviously, from Eq. \eqref{eq:theta_mon_gauge}, this requires that $M_1 \left( r \right) = 0$. Substituting $M_1 \left( r \right) = 0$ back into $\delta h_{ab}$ leaves us with a $tr$-component which can vanish only if $M_0 ' \left( r \right)= 0$. The remaining gauge degree of freedom $M_0$ must therefore be a constant. However, the condition $h_{tt} = \delta h_{tt}$ leads to $M_0 \left( r \right) = - i H_0 \left( r \right) / 2 \omega$. This shows that monopolar $tt$-only-gauge perturbations must have a constant $H_0 \left( r \right)$, and this special class excludes the hairy perturbations given by Eqs. \eqref{eq:monopole_hair_tfl} and \eqref{eq:monopole_hair_stealth}.

\subsection{$tt$-only-dipole gauge mode}
\label{subsec:tt_dipole}

For the even parity dipole, $l = 1$, we focus on $m = 0$ for practical calculations noting that all three dipole modes, $m = 0, \pm 1$, can be rotated into each other owing to the spherical symmetry of the background.
We warn, however, that even with such simplification this case remains to be more technically involved compared to the monopolar counterpart discussed previously. Moving on, in this case, the gauge vector $\xi^a$ can be decomposed as
\begin{equation}
\xi^a = \left( M_0 \left( r\right) Y_{10} e^{-i \omega t}, M_1 \left( r\right) Y_{10} e^{-i \omega t} , M_2 \left( r \right) \left( \partial_\theta Y_{10} \right) e^{-i \omega t}, 0 \right) .
\end{equation}
With this, the gauge transformed metric, $\delta h_{ab} = 2\nabla_{(a} \xi_{b)}$, has the following independent components:
\begin{eqnarray}
\label{eq:tt_dip_gauge} \delta h_{tt} &=& \frac{1}{2} i \sqrt{\frac{3}{\pi }}  \cos (\theta ) e^{-i t \omega } \left(2 \omega  f M_0+i M_1 f' \right) \\ 
\label{eq:tr_dip_gauge} \delta h_{tr} &=& - \dfrac{1}{2}\sqrt{\frac{3}{\pi }} \cos (\theta ) e^{-i t \omega } \frac{ \left(f^2 M_0'+i \omega  M_1\right)}{ f } \\
\label{eq:rr_dip_gauge} \delta h_{rr} &=& \frac{1}{2} \sqrt{\frac{3}{\pi }}  \cos (\theta ) e^{-i t \omega }  \frac{\left(2 f M_1'-M_1 f'\right)}{ f^2} \\
\label{eq:t_theta_dip_gauge} \delta h_{t\theta} &=& \frac{1}{2} \sqrt{\frac{3}{\pi }}  \sin (\theta ) e^{-i t \omega } \left(f M_0+i r^2 \omega  M_2 \right) \\
\label{eq:r_theta_dip_gauge} \delta h_{t\theta} &=& - \frac{1}{2}\sqrt{\frac{3}{\pi }} \sin (\theta ) e^{-i t \omega } \frac{ \left(r^2 f M_2'+M_1\right)}{ f} \\
\label{eq:theta_dip_gauge} \delta h_{\theta\theta} &=& - \sqrt{\frac{3}{\pi }} r \cos (\theta ) e^{-i t \omega } (r M_2 - M_1) .
\end{eqnarray}
Metric perturbations $h_{ab}$ which can be accommodated instead by using the above components of gauge transformed metric perturbation $\delta h_{ab}$ are gauge modes. This includes the center of mass-dipolar gauge mode discussed in detail by Zerilli \cite{zerilli_classic_2}.

Focusing on $tt$-only-gauge modes, we search for the gauge vector $\left( M_0, M_1, M_2 \right)$ for which the resulting metric perturbation $\delta h_{ab}$ has only a nonzero $tt$-component which we write down as
\begin{equation}
h_{tt} = \frac{1}{2} \sqrt{\frac{3}{\pi }}  f H_0 \cos (\theta ) e^{-i t \omega } .
\end{equation}
In line with this goal, Eq. \eqref{eq:theta_dip_gauge} shows that $M_2 = M_1/r$, thus, spending our first gauge degree of freedom. Now, solving for $M_1$ in $h_{tt} = \delta h_{tt}$ we obtain
\begin{equation}
M_1 = - \dfrac{- f H_0 + 2 i \omega f M_0}{ f' }  ,
\end{equation}
leaving us with one remaining gauge degree of freedom in $M_0$. We spend this by making the $t\theta$-component of $\delta h_{ab}$ vanish and obtain
\begin{equation}
M_0 = - \dfrac{ i r \omega H_0 }{ 2 r \omega^2 - f' } .
\end{equation}
At this point, we have the desired form of the $tt$-component already, but this still comes with the remaining nonzero $tr$-, $rr$-, and $r\theta$-components which can be written as
\begin{equation}
\label{eq:tr_boss_dip}
H_0'=\frac{ H_0 \left(r \left(2 \omega ^2-f f''\right)+(f-1) f'\right)}{r f \left(2 r \omega ^2-f'\right)} ,
\end{equation}
\begin{equation}
\label{eq:rr_boss_dip}
H_0'=\frac{H_0 \left(f \left(4 \omega ^2-2 f''\right)+f' \left(f'-2 r \omega ^2\right)\right)}{2 f \left(2 r \omega ^2-f'\right)} ,
\end{equation}
and
\begin{equation}
\label{eq:rtheta_boss_dip}
H_0'=-\frac{H_0 \left(\left(r f'+1\right) \left(2 r \omega ^2-f'\right)+f \left(r \left(f''-4 \omega ^2\right)+f'\right)\right)}{r f \left(2 r \omega ^2-f'\right)} ,
\end{equation}
respectively. The common solution $H_0$ to the three equations shown above describes the class of $tt$-only even-parity dipolar gauge modes. However, by equating $H_0'$ from any of the above equations into the other two, one would find that a necessary condition for a solution of all three equations is given by $r f' - 2 f + 2 = 0$. This restricts the form of the metric function to $f = 1 + c r^2$, where $c$ is an integration constant, for which nontrivial $tt$-only even-parity dipolar gauge modes can be found. Even though this is a subset of the Schwarzschild-(A)dS family of $f(r)$, the missing ``mass" term, $-2M/r$, shows that in general the remaining differential equations cannot be simultaneously satisfied. Hence, the $tt$-only-perturbation for the stealth Schwarzschild-(anti) de Sitter black holes in KGB given by Eqs. \eqref{eq:dipole_hair_tfl} and \eqref{eq:dipole_hair_stealth} cannot be gauged away, in general. 

In the uninteresting case of $M = 0$, with $f = 1 - \left( \Lambda r^2/3 \right)$ and $\Lambda = -\lambda/2\kappa$, Eqs. \eqref{eq:tr_boss_dip}, \eqref{eq:rr_boss_dip}, and \eqref{eq:rtheta_boss_dip} reduce to the same equation given by $ H_0' = 6 \kappa H_0 / \left( 6 r \kappa + r^3 \lambda \right) $. The exact solution to this is given by $H_0 = c r / \sqrt{ 6 \kappa + \lambda r^2 }$ where $c$ is an integration constant.

We end by specializing the above discussion for the Schwarzschild case. Substituting $f = 1 - 2M/r$ into Eqs. \eqref{eq:tr_boss_dip}, \eqref{eq:rr_boss_dip}, and \eqref{eq:rtheta_boss_dip} and solving for $H_0$ lead to
\begin{equation}
H_0 = c \dfrac{ \left( r - 2M \right) \left( M - r^3 \omega^2 \right) }{r^3} ,
\end{equation}
\begin{equation}
H_0 = c \dfrac{ M - r^3 \omega^2 }{ r^{3/2} \sqrt{ r - 2M } } ,
\end{equation}
and
\begin{equation}
H_0 = c \dfrac{ M - r^3 \omega^2 }{ \left( r - 2M \right)^2 } ,
\end{equation}
respectively. The above solutions can be equal only in the trivial case $c = 0$, unless $M = 0$. In the context of Section \ref{subsec:even_kgb}, the desired $H_0$ is given by Eq. \eqref{eq:dipole_hair_stealth} which in the Schwarzschild case becomes equal to $H_0 = r^{3/2} q\left( \omega \right) / \left( 3 M \kappa \sqrt{r - 2M} \right)$. This clearly cannot be a gauge mode.

\section{Generic structure of the cones in KGB}
\label{sec:cones_kgb}

It is possible to derive an analytic expression for the scalar cone in KGB and through this support the conclusion that the scalar field becomes nondynamical on a stealth background. Starting with the generic perturbations $\left( h_{ab}, \psi \right)$ in the action \eqref{eq:kgb}, performing a linear transformation of the metric perturbation,
\begin{equation}
\tilde{h}_{ab} = h_{ab} - \dfrac{h}{2} g_{ab} - \dfrac{ G_X }{ \kappa } \phi_a \phi_b \psi ,
\end{equation}
and then imposing the transverse-gauge condition on $\tilde{h}_{ab}$, i.e., $\nabla_{b} \tilde{h}^{ab} = 0$, 
it can be shown that the linearized field equations for $\left( \tilde{h}_{ab}, \psi \right)$ reduce to
\begin{equation}
\Box \tilde{h}_{ab} + \mathcal{O} \left( \nabla \tilde{h}, \nabla \psi  \right) = 0
\end{equation}
and 
\begin{equation}
\mathcal{G}^{ab} \psi_{ab} + \mathcal{O} \left( \nabla \tilde{h}, \nabla \psi  \right) = 0
\end{equation}
where $\mathcal{O} \left( \nabla \tilde{h}, \nabla \psi  \right)$ are terms with less than two spacetime derivatives of the perturbations \cite{gw_birefringence_ezquiaga}. See the supplementary Mathematica notebook \textit{cones\_of\_kgb.nb} in the author's website for a detailed derivation. The above expression implies that the tensor modes, i.e., gravitational waves in KGB, propagate on the light cone and the scalar perturbation on the other hand move on the cone defined by the effective metric
\begin{equation}
\label{eq:scalar_cone_kgb}
\begin{split}
\mathcal{G}^{ab} =  A g^{ab} + B \phi^a \phi^b
\end{split}
\end{equation}
where $A$ and $B$ are functionals of the potentials on the background and given by
\begin{eqnarray}
A &=& K_X - 2 G_X \Box \phi - \dfrac{X^2}{\kappa} G_X^2 - \phi^c X_c G_{XX} - 2 G_\phi + 2 X G_{\phi X} \\
B &=& - K_{XX} - \dfrac{2X}{\kappa} G_X^2 + G_{XX} \Box \phi + 2 G_{\phi X}
\end{eqnarray}
and $X_a = \nabla_a X = - \phi^b \phi_{ab}$. This result can be used as starting point of a gravitational wave analysis in any background in KGB. 

In a spherically symmetric background,
\begin{eqnarray}
ds^2 &=& - h\left( r \right) dt^2 + \dfrac{dr^2}{f\left(r\right)} + r^2 d\Omega^2 \\
\phi &=& \phi \left( r \right) ,
\end{eqnarray}
we can use this to write down the generic scalar mode equation
\begin{equation}
\left( A + B f \phi^{\prime 2} \right) \left(- \dfrac{\partial_t^2 \psi}{c_s^2 h} + f \partial_r^2 \psi \right) + \mathcal{O} \left( \nabla_{\theta , \varphi} \tilde{h}, \nabla_{\theta , \varphi} \psi  \right) = 0
\end{equation}
where the sound speed $c_s$ is given by
\begin{equation}
c_s^2 = 1 + \dfrac{B}{A} f \phi^{\prime 2} .
\end{equation}
This shows for example that in quintessence, $K \left( \phi, X \right) = X - V\left( \phi \right)$, the sound speed reduces to unity. Most importantly, this provides an independent calculation confirming that for stealth black holes in shift symmetric theory, defined by $K_X = 0$ and $G_X = 0$ which translates to $A = 0$ and $B \neq 0$, the sound speed becomes infinite. The scalar cone therefore opens up and the scalar field can no longer be considered a propagating degree of freedom.

%\bibliographystyle{JHEP}
%\bibliography{bibfile}

\providecommand{\href}[2]{#2}\begingroup\raggedright\endgroup

\end{document}